\documentclass[sigchi]{acmart}

\usepackage{booktabs} 

\usepackage{makecell}
\usepackage{dblfloatfix}

\usepackage{subcaption}
\usepackage{balance}

\usepackage{tikz}
\usepackage{tikzpagenodes}

\usepackage{lipsum}
\usepackage{enumitem}

\copyrightyear{2019}
\acmYear{2019}
\setcopyright{acmlicensed}
\acmConference[CHI 2019]{CHI Conference on Human Factors in Computing Systems Proceedings}{May 4--9, 2019}{Glasgow, Scotland UK}
\acmBooktitle{CHI Conference on Human Factors in Computing Systems Proceedings (CHI 2019), May 4--9, 2019, Glasgow, Scotland UK}
\acmPrice{15.00}
\acmDOI{10.1145/3290605.3300830}
\acmISBN{978-1-4503-5970-2/19/05}

\usepackage{xspace}

\begin{document}
\title{Improving Fairness in Machine Learning Systems:\\ What Do Industry Practitioners Need?}
\renewcommand{\shorttitle}{Improving Fairness in Machine Learning Systems}

\author{Kenneth Holstein}
\affiliation{%
  \institution{Carnegie Mellon University}
  \city{Pittsburgh}
  \state{PA}
}
\email{kjholste@cs.cmu.edu}

 \author{Jennifer Wortman Vaughan}
\affiliation{%
  \institution{Microsoft Research}
  \city{New York}
  \state{NY}
}
\email{jenn@microsoft.com}

\author{Hal Daum{\'e} III}
\affiliation{%
  \institution{Microsoft Research \& \\ University of Maryland}
  \city{New York}
  \state{NY}
 }
\email{me@hal3.name}

\author{Miroslav Dud{\'i}k}
\affiliation{%
  \institution{Microsoft Research}
  \city{New York}
  \state{NY}
 }
  \email{mdudik@microsoft.com}

\author{Hanna Wallach}
\affiliation{%
  \institution{Microsoft Research}
  \city{New York}
  \state{NY}
 }
 \email{wallach@microsoft.com}

\renewcommand{\shortauthors}{K. Holstein et al.}
\newcommand{\includesection}[2]{\section{#2} \label{sec:#1} \input{#1}}
\newcommand{\includesubsection}[2]{\subsection{#2} \label{sec:#1} \input{#1}}

\def\equationautorefname~#1\null{Eq~#1\null}
\renewcommand{\sectionautorefname}{\S\kern-0.2em}
\renewcommand{\subsectionautorefname}{\S\kern-0.2em}
\renewcommand{\subsubsectionautorefname}{\S\kern-0.2em}

\definecolor{cerulean}{rgb}{0.10, 0.58, 0.75}
\newcommand{\mdedit}[1]{\textcolor{cerulean}{#1}\xspace}

\newcommand{\inlinecomment}[3]{\textcolor{#2}{[\textsf{\textbf{#1: #3}}]}}
\newcommand{\hal}[1]{\inlinecomment{hal}{orange!80!white}{#1}}
\newcommand{\ken}[1]{\inlinecomment{ken}{violet!80!white}{#1}}
\newcommand{\jenn}[1]{\inlinecomment{jenn}{purple!80!white}{#1}}
\newcommand{\hanna}[1]{\inlinecomment{hanna}{olive!80!white}{#1}}
\newcommand{\miro}[1]{\inlinecomment{miro}{cerulean}{#1}}

\renewenvironment{quote}%
  {\list{}{\leftmargin=0.2in\rightmargin=0.2in}\item[]}%
  {\endlist}

\begin{abstract}
The potential for machine learning (ML) systems to amplify social
inequities and unfairness is receiving increasing popular and academic
attention. A surge of recent work has focused on the development of algorithmic tools to
assess and mitigate such unfairness. If these tools are to have a
positive impact on industry practice, however, it is crucial that
their design be informed by an understanding of real-world
needs. Through 35 semi-structured interviews and an anonymous survey
of 267 ML practitioners, we conduct the first systematic investigation
of commercial product teams' challenges and needs for support in
developing fairer ML systems. We identify areas of alignment and
disconnect between the challenges faced by teams in practice and
the solutions proposed in the fair ML research literature. Based on these
findings, we highlight directions for future ML and HCI research that
will better address practitioners' needs.

\end{abstract}

%
%

\ccsdesc[500]{Human-centered computing}
\ccsdesc[500]{Social and professional topics~Socio-technical systems}
\ccsdesc[500]{Computing methodologies~Machine learning}

\keywords{algorithmic bias, fair machine learning, product teams, need-finding, empirical study, UX of machine learning}

\maketitle

\section{Introduction} \label{sec:intro} 

Machine learning (ML) systems increasingly influence every facet of our lives, including the quality of healthcare and education
we receive
\cite{ferryman2018fairness,esteva2017dermatologist,holstein2018student,bosch2016detecting},
which news or social media posts we see
\cite{rader2015understanding,alvarado2018towards,bucher2017algorithmic},
who receives a job
\cite{hirevue,pymetrics},
  who is released from jail~\cite{angwin2016machine,chouldechova2017fair},
  and who is subjected to increased policing~\cite{lum2016predict,larson2016we,veale2018fairness}.  With this growth, the potential of ML to amplify social inequities has received growing attention across several research communities, as well as in the popular press. It is now commonplace to see reports in mainstream media of systemic unfair behaviors observed
in widely used ML systems---for example, an automated hiring system that is more likely to recommend hires from certain racial, gender, or age groups~\cite{wachter-boettcher_2017,giang_2018}, or a search engine that amplifies negative 
stereotypes by 
showing arrest-record ads in response to queries for
names predominantly given to African American babies, but not for other names~\cite{bbc_news_2013,noble2018algorithms}.

Substantial effort in the rapidly growing research literature on fairness in ML
has centered on the development of statistical definitions of fairness~\cite{21definitions,chouldechova2017fair,dwork2012fairness,hardt2016equality} 
and algorithmic methods to assess and mitigate undesirable biases in relation to these definitions~\cite{agarwal2018reductions,hardt2016equality,kusner2017counterfactual}. 
As the field matures, integrated toolkits are 
being developed with the aim of making these 
methods more widely accessible and usable
(e.g., \cite{galhotra2017fairness,aequitas,lomas_2018,google_whatif_2018,angell2018themis,lomas_2018_IBM}).
While some fair ML 
tools are already being prototyped with practitioners, their initial design often appears to be driven more by the availability of algorithmic methods than by real-world needs (cf.~\cite{holstein2017intelligent, yang2016investigating}).
If such tools are to have a positive and meaningful impact on industry practice, however, it is crucial that their design 
be informed by an understanding of practitioners' actual challenges and needs for support in developing fairer ML systems~\cite{yang2016investigating}.\looseness=-1

We investigate the challenges faced by
commercial ML product teams---whose products affect the lives of millions of users~\cite{buolamwini2018gender, springerGarciaCramer2018}---in monitoring for unfairness 
and taking appropriate action~\cite{springerGarciaCramer2018, veale2018fairness}. Through semi-structured interviews with 35 practitioners, across 25 ML product teams from 10 major companies, we investigate teams' existing practices and challenges around fairness in ML, as well as their needs for additional support. To better understand the prevalence and generality of the key themes surfaced in 
our interviews, we then conduct an anonymous survey of 267 industry ML practitioners, across a broader range of contexts. 
To our knowledge, this is the first systematic investigation of industry ML practitioners' challenges and needs around fairness.\looseness=-1

Through our investigation, we identify a range of real-world needs that have been neglected in the literature so far, as well as several areas of alignment.
For example, while the fair ML literature has largely focused on ``de-biasing'' methods and viewed the training data as fixed~\cite{chen2018my,kallus2018residual}, most of our interviewees report that their teams consider data collection, rather than model development, as the most important place to intervene.
Participants also often report struggling to apply existing auditing and de-biasing methods in their contexts.
For instance, whereas previously proposed 
methods typically require access to sensitive demographics at an individual level, such information is
frequently available only at coarser levels.
The fair ML literature has tended to focus on domains such as recidivism prediction, automated hiring, and face recognition, where
fairness can be understood, at least partially, in terms of well-defined quantitative metrics
~\cite{chouldechova2018case,kleinberg2016inherent,buolamwini2018gender}, whereas teams working on applications involving richer interactions between the user and the system
(e.g., chatbots, web search, and adaptive tutoring) brought up needs for more holistic, system-level fairness auditing methods.
Interviewees also stressed the importance of explicitly considering biases and ``blind spots'' that may be present in the humans embedded throughout the ML development pipeline, such as crowdworkers 
or user-study participants. 
Such concerns also extended to product teams' own blind spots. For instance, teams often struggled to anticipate
which subpopulations and forms of unfairness
they need to consider for specific kinds of ML applications.

Based on these findings, we highlight opportunities for the ML and HCI research communities to have a greater impact on industry and on the fairness of ML systems in practice.


\section{Background and Related Work} \label{sec:background} 

The design, prototyping, and maintenance of machine learning systems raises many unique challenges~\cite{dove2017ux,kulesza2015principles,nushi2017human,sculley2015hidden}
not commonly faced with other kinds of intelligent systems or computing systems more broadly~\cite{friedman1996bias,lee2018understanding,lee2017algorithmic}.
The budding area of ``UX for ML'' has begun to explore 
new forms of prototyping for ML systems, to provide earlier insights into the complex, interacting UX impacts of particular dataset,
modeling, and system-design choices~\cite{dove2017ux,googleDesign,holstein2018classroom,yang2018machine}. In addition, a growing body of research focuses on the design and development
of programmer tools that can better support developers in debugging and effectively monitoring the predictive performance of complex ML systems~\cite{kulesza2015principles,chen2018anchorviz,amershi2015modeltracker,nushi2017human,kery2018story}.

In parallel, the potential for undesirable biases in ML systems to exacerbate existing social inequities---or even generate new ones---has received considerable attention across a range of academic disciplines, from ML to HCI to public policy, law, and ethics~\cite{springerGarciaCramer2018,veale2018fairness,barocas2016big}.
Specialized research conferences and initiatives are forming with a focus on biases and unfairness in data-driven
algorithmic systems, such as the Workshop on Fairness, Accountability, and Transparency in Machine Learning (FAT/ML)~\cite{FATML}, the nascent FAT* conference~\cite{FAT*}, AI Now~\cite{AINow}, and the Partnership on AI~\cite{partnershipOnAI}.
Significant effort in the fair ML community has focused on the development of statistical definitions of fairness~\cite{21definitions,chouldechova2017fair,berk2017fairness,dwork2012fairness} 
and algorithmic methods to assess and mitigate biases in relation to these definitions~\cite{agarwal2018reductions,hardt2016equality,kusner2017counterfactual,bolukbasi2016man}.
In contrast, the HCI community has studied unfairness in ML systems through political, social, and psychological lenses, among others (e.g., \cite{binns2017fairnesspoliticalphilosophy,hamidi2018gender,schlesinger2018let,woodruff2018qualitative}). For example, HCI researchers have empirically studied users' expectations and perceptions related to fairness in 
algorithmic systems, finding that these do not always align with existing statistical definitions~\cite{lee2018understanding,lee2017algorithmic, binns2018s,woodruff2018qualitative}.
Other 
work has focused on auditing widely-used ML products from the outside~\cite{angwin2016machine, buolamwini2018gender,kay2015unequal,diaz2018addressing}, and often concluded with high-level calls to action aimed at those responsible for developing and maintaining these systems or for regulating their use~\cite{reisman2018algorithmic, AAprimer,WWWfoundation}. Lastly, \citet{crawford_2017}, \citet{springerGarciaCramer2018}, and others have highlighted an urgent need for
internal processes and tools to support companies in developing fairer systems in the first place.

Despite this widespread attention to biases and unfairness in ML, to the best of our knowledge, only one prior study, by \citet{veale2018fairness}, has investigated actual ML practitioners' challenges and needs for support in creating fairer ML systems.
Veale et al.\ conducted exploratory interviews with public-sector ML practitioners working across a range of high-stakes contexts, such as predictive policing~\cite{lum2016predict,larson2016we}
and child mistreatment detection~\cite{chouldechova2018case},
to understand
the
challenges faced by practitioners in aligning the behavior of ML systems with public values.
Through these interviews, the authors uncovered
several disconnects between the real-world challenges that arise in public-sector ML practice compared with those commonly presumed in the fair ML literature.

In the same spirit as \citet{veale2018fairness}, this work investigates ML practitioners' needs for support, with the aim of identifying fruitful opportunities for future research \cite{stokes2011pasteur}.  Whereas Veale et al.\ studied algorithm-assisted
decision makers in high-stakes public-sector contexts---who are often experienced in thinking about fairness, yet relatively new to working with ML systems---we study industry ML practitioners, who tend to be experienced in developing ML systems, but relatively new to thinking about fairness.  Supporting industry ML practitioners can also be viewed as a critical step towards fairer algorithm-assisted decision making downstream, including in the public sector, where systems are often built on top of ML products and APIs developed in industry~\cite{gershgorn_2018c,buolamwini2018gender,thubron_2018,powles2017google}.
Unlike the public-sector practitioners studied by Veale et al., the industry practitioners studied here work on a much broader range of applications, such as image captioning, web search, chatbots, speech recognition, and  personalized retail. In many of these applications, ethical considerations may be even less clear cut than in high-stakes public-sector contexts. Moreover, motivations and organizational priorities can differ considerably in industry contexts.


\section{Methods} \label{sec:method} 

\renewcommand{\arraystretch}{0.70}
\begin{table*}[t]
  \caption{Interview demographics: Interviewees' self-reported technology areas and team roles. Where multiple people were interviewed from the same product team, interviewee identifiers are grouped in square brackets.
}
  \label{tab:freq}
\small
  \centering
  \def\arraystretch{1.1}
  \begin{tabular}{p{1.7in}@{~~~~}p{3.4in}@{~~~~}p{1.6in}}
    \toprule
    Technology Area &Roles of Interviewees & Interviewee IDs \\
    \midrule
    Adaptive Tutoring \& Mentoring
                    & Chief Data Scientist, CTO, Data Scientist, Research Scientist
                    & R10, {[}R13, R14{]}, R30 \\
    Chatbots
                    & CEO, Product Mgr., UX Researcher
                    & {[}R17, R18{]}, R35 \\
    Vision \& Multimodal Sensing
                    & CTO, ML Engineer, Product Mgr., Software Engineer
                    & {[}R2, R3, R4{]}, R6, R7, R9, R26 \\
    General-purpose ML (e.g., APIs)
                    & Chief Architect, Director of ML, Product Mgr.
                    & R25, R32, R34\\
    NLP (e.g., Speech, Translation)
                    & Data Mgr., Data Collector, Domain Expert, ML Engineer, Prod- & R1, {[}R15, R16, R19, R20, R21, \\ & uct Mgr., Research Software Eng., Technical Mgr., UX Designer
                    & R22{]}, R24, {[}R27, R29{]}, R28, R31 \\
    Recommender Systems 
                    & Chief Data Scientist, Data Scientist, Head of Diversity Analytics
                    & R8, R12, R23, R33 \\
    Web Search & Product Mgr. & R5, R11\\
  \bottomrule
  \end{tabular}
\end{table*}

To better understand product teams' 
needs for support
in developing fairer ML systems, we conducted a series of semi-structured, one-on-one interviews with a total of 35 practitioners, across 25 ML product teams from 10 major companies. 
To investigate the prevalence and generality of the key themes that emerged in these interviews, we then conducted an anonymous survey with a broader sample of 267 industry ML practitioners.
For both the interviews and the survey, ``practitioners'' were defined broadly as those who work in any role on a team that develops products or services involving ML.
The study went through an ethical review and was IRB-approved.
Semi-structured interview protocols and survey questions are provided in the supplementary materials.\looseness=-1


\subsection{Interview Study} \label{sec:method-interview} 

Our interview study proceeded in two rounds. First, to get a broad sense of challenges and needs, we conducted six formative interviews.
Building on emerging themes,
we then conducted more
in-depth
interviews with a larger sample.\looseness=-1

The first six interviews were conducted with product managers (PMs) across different technology areas. Each interview lasted 30 minutes and was conducted by teleconference because PMs were distributed across multiple countries.
Each PM was first asked to describe the products their team is responsible for, who the customers of these products are, and how their team is structured. Interviewees were then asked whether fairness is something their team regularly discusses or incorporates into their workflow. The meaning of ``fairness'' was intentionally left open as we were interested in hearing PMs' notions about what it might mean for their products to be fair. However, if a PM requested clarification at any point, we provided a broad definition: \emph{``Any case where AI/ML systems perform differently for different groups in ways that may be considered undesirable.''}
PMs were then asked whether their team or customers had ever encountered issues relating to fairness in their products. If so, PMs were asked for concrete examples, 
and were asked high-level follow-up questions about these experiences.
Otherwise,
they were asked whether they thought such issues might exist undetected, and whether they had seen \emph{``other surprising or unexpected issues arise''} (cf.~\cite{veale2018fairness}). These follow-ups sometimes led PMs to realize they actually did have relevant experiences to share.\looseness=-1

Our second, main round of interviews built on themes that emerged during the initial round. We conducted more detailed, semi-structured interviews to investigate teams' current practices, challenges, and needs in greater depth. Interviewees included 29 practitioners, across 19 ML product teams from 10 major technology companies.
As shown in \autoref{tab:freq}, we interviewed practitioners across a range of
technology areas and team roles.
Whenever possible, we tried to interview people in different roles on the same team to hear (potentially) different perspectives. 
Interviewees were recruited using snowball sampling.
We searched for news articles related to ML biases and unfairness and contacted members of teams whose products had previously received relevant media coverage (i.e., news articles about unfair behavior observed in these products).
In addition, we emailed direct contacts across over 30 major companies. In both cases, we asked contacts to share our interview invitation with any colleagues working on ML products (in any role) at their own company or others. At the end of each interview, interviewees were again encouraged to share any
relevant contacts.\looseness=-1

Although prospective interviewees were often eager to participate and recruit colleagues, we encountered several challenges resembling those discussed by Veale et al.~\cite{veale2018fairness}.
For instance, given a recent trend of negative media coverage calling out algorithmic biases and unfairness in widely-used ML systems (e.g., \cite{bbc_news_2013,thubron_2018,wachter-boettcher_2017,zhang_2015}), our contacts often expressed strong fears that
their team or company's identity
might leak to the popular press, harming their reputation. Some contacts revealed a general distrust of researchers, citing cases where researchers have benefited by publicly critiquing companies' products from the outside instead of engaging 
to help them improve their products. Finally, some contacts worried that, in diving into the details of their teams' prior experiences, 
they might inadvertently reveal trade secrets. 
For these reasons, contacts often declined to be interviewed. 
To allay some of these concerns, we assured contacts that the goal of these interviews was to help us learn about teams' current practices, challenges, and needs around fair ML in general, and that findings would not be linked to specific individuals, teams, or companies. We also asked for interviewees' advance permission to audio record the interviews, noting that these recordings would not be shared outside of the research team. Furthermore, we noted that these
recordings would be destroyed following transcription, and that the resulting
transcriptions would be de-identified. Finally, we assured interviewees that we would allow them to review any (de-identified) direct quotes 
before including them in any research publications. All 29 interviewees in the main round of interviews consented to be audio recorded.\looseness=-1

Each interview in the main round lasted between 40 and 60 minutes. In each one, interviewees were first reminded of the overall purpose of the research, and were then asked a series of questions about fairness at each stage in their team's ML development pipeline---from collecting data to designing datasets (e.g., curating training and test sets) and developing an ML product to assessing and potentially mitigating fairness issues in that product.
For each of these stages, interviewees were asked a broad opening question about critical episodes they had encountered 
(e.g., \emph{``Can you recall times you or your team have discovered fairness issues in your products?''}), and a series of follow-up questions (where applicable and not previously covered). While a few follow-up questions were specific to particular stages of the pipeline, a core sequence of four follow-ups was used across all stages. First, interviewees were asked to walk through how their team navigated the episode (e.g., \emph{``When you decided there were issues that needed to be addressed... how did your team decide what course of action to take to address them?''}). Next, interviewees were instructed to imagine they could return to these critical episodes, but this time with access to a magical oracle of which they could ask any question to help them in the moment~\cite{holstein2017intelligent,kery2018story}. We asked the question in this way to encourage interviewees to speak freely about their
current challenges and needs 
without feeling constrained to those for which they believed a solution was currently 
available~\cite{holstein2017intelligent,kery2018story}. After interviewees generated questions for the oracle, they were then asked, for each question, whether and how their team currently goes about trying to answer this question.
This follow-up often led interviewees to reflect on gaps between their current and ideal practices. Finally, interviewees were asked whether they saw any other areas for improvement, or opportunities for support, related to the relevant stage of their team's ML development pipeline.

\begin{figure*}
\hspace*{0mm}
\fbox{\includegraphics[trim={0cm 0 0.25cm 0},clip,height=0.29\textwidth,keepaspectratio]{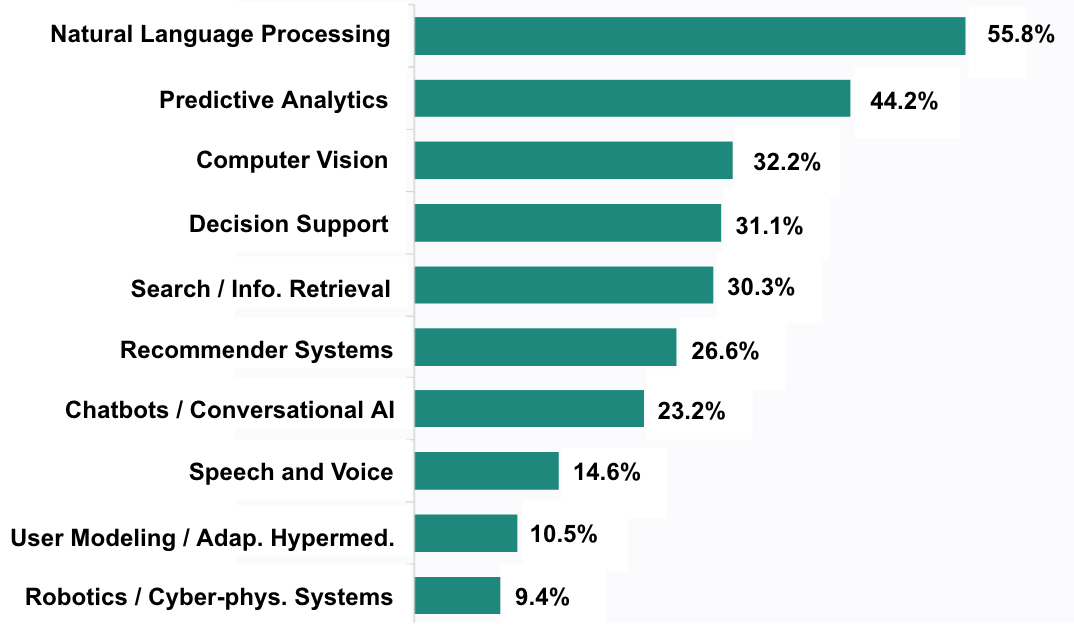}}
\hspace*{0mm}
\fbox{\includegraphics[trim={0.1cm 0 0.25cm 0},clip,height=0.29\textwidth,keepaspectratio]{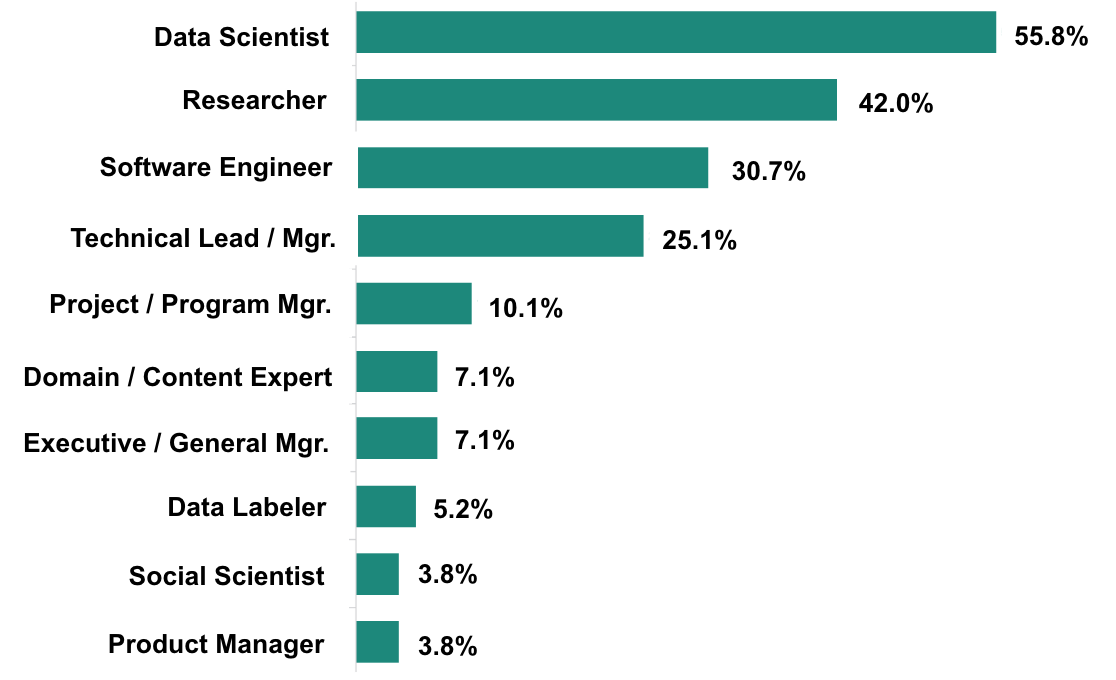}}
\caption{
Survey demographics: the top 10 self-reported technology areas (left) and team roles (right).
}
\label{fig:demo}
\end{figure*}

To analyze the interview data, we worked through
transcriptions of approximately 25 hours of audio 
to synthesize findings using standard methodology
from contextual design: interpretation sessions and affinity diagramming~\cite{holtzblatt1993contextual,hanington2012universal}. Specifically, we employed bottom-up affinity diagramming (using MURAL~\cite{mural}) to iteratively generate codes for various interviewee utterances and then grouped these codes into successively higher-level themes concerning current
practices,
challenges, and needs for support. Key themes are presented in detail below, under \textit{Results and Discussion}.\looseness=-1

%


\subsection{Survey} \label{sec:method-survey} 

To validate our findings on a broader population, we conducted an anonymous online survey using Qualtrics~\cite{qualtrics2013qualtrics}. Respondents were recruited using snowball sampling. Specifically, we emailed the survey to direct contacts at over 40 companies and invited them to pass the survey on to colleagues (within or outside of their companies) who are part of a team working on ML products (in any role). We also announced the survey on social media (e.g., Twitter) and online communities related to ML and AI, including Kaggle forums and special interest groups on LinkedIn, Facebook, Reddit, and Slack. 
\looseness=-1

We structured the survey to act as a quantitative supplement to the interviews. As such, the high-level structure of the survey mirrored that of the main round of interviews. Based on our findings from the interviews, we developed survey questions to investigate the prevalence and generality of emerging themes. 
First, we asked a set of demographic questions to understand our respondents' backgrounds, including their technology area(s) and team role(s). In a branching sequence of survey sections, respondents were then asked about their team's current practices, challenges, and needs for support around fairness, with each section pertaining to one stage of their team's ML development pipeline. For each question, closed-ended response options were provided based on themes that emerged from the interviews through affinity diagramming, in addition to free-response options that allowed respondents to elaborate on their responses.

A total of 287 people started the survey. However, not all respondents
completed it,
so we analyze only the
267
respondents who completed at least one section beyond demographics. 
The top 10 self-reported
technology areas and team roles are shown in Figure~\ref{fig:demo}. In each case, respondents were able to select multiple options. Additional survey
demographics are available in the supplementary materials.


\section{Results and Discussion} \label{sec:results} 

Although participants spanned a diverse range of companies, team roles, and application areas,
we observed many commonalities. In the following, we discuss teams' current 
challenges and needs around fairness, organized by top-level themes that emerged
through affinity diagramming.
These include needs for support in fairness-aware data collection and curation, 
overcoming teams' blind spots, 
implementing more proactive fairness auditing processes, 
auditing complex ML systems, 
deciding how to address particular instances of unfairness, 
and addressing biases in the humans embedded throughout the ML development pipeline.
Within each of these top-level themes, we present selected sub-themes to highlight research and design opportunities that have received little attention in the fair ML literature thus far.

We supplement our findings from the interviews with corresponding survey responses. Interviewees are identified with an ``R,'' and survey responses are accompanied with percentages. 
Some survey questions were completed by a subset of respondents because the survey used branching logic, with follow-up sections appearing only as applicable (e.g., respondents were only asked questions about addressing fairness issues if they reported that their team had previously detected such issues in their products). In such cases, question-specific response rates 
are provided in addition to the percentage of respondents who were asked the question. A graphical summary of selected survey responses is provided in the supplementary materials.  To illustrate general themes, we share direct quotes in cases where we received explicit permission from interviewees, in accordance with our IRB approval and consent form. Therefore, although the themes we present are drawn from a wide range of application domains, the domains represented by direct quotes may be narrower.\looseness=-1

By recruiting interviewees using snowball sampling and specifically targeting members of teams whose products had previously received media coverage related to ML biases and unfairness, we may have sampled practitioners who are unusually motivated to address fairness issues in their products. Indeed, many interviewees reported having invested significant time and effort into trying to improve fairness in their teams' products, even when they felt unsupported in these initiatives by their team or company leadership. As such, many of the findings presented below may be taken to represent barriers that industry ML practitioners run up against in improving fairness even when they are motivated to do so. \looseness=-1

While much of our discussion focuses on needs for tools, we stress that biases and unfairness are fundamentally socio-technical problems, and that technically focused research efforts must go hand-in-hand with efforts to improve organizational processes, policies, and education \cite{selbst2018fairness, toyama2018needs}.


\subsection{Fairness-aware Data Collection} \label{sec:results-collection} 

While media coverage of biases and unfairness in ML systems often uses a ``bias in, bias out'' framing, emphasizing the central role of dataset quality \cite{gershgorn_2018b,wachter-boettcher_2017}, the fair ML research literature has overwhelmingly
focused on the development of algorithmic methods to mitigate biases,
viewing the dataset as fixed 
\cite{chen2018my,kallus2018residual}.
However, many of our interviewees reported that their teams typically look to their training datasets, not their ML models, as the most important place to intervene to improve fairness in their products.
Out of the 65\% of
survey respondents
who reported that their teams have some control over data collection and curation, a majority (58\%) reported that they currently consider fairness at these stages of their ML development pipeline.
Furthermore, out of the 21\% of
respondents whose teams had previously tried to address fairness issues found
in their products, the most commonly attempted strategy (73\%) 
was ``collecting more training data.''\looseness=-1

\subsubsection{"It's almost like the wild west."\nopunct}
Most interviewees reported that their teams do not currently have processes in place to support the collection and curation of balanced or representative datasets. 
A software engineer (R7) described their team's current data collection practices as \emph{``almost like the wild west''} and a data scientist (R10) noted that \emph{``there isn't really a thought process surrounding... `Should [our team] ingest this data in?' [...] If it is available to us, we ingest it.''}
An ML engineer (R19) emphasized the value of supporting 
more effective communication (e.g., around sampling practices and useful metadata to include) between 
those responsible for decisions about data collection and curation, and 
those responsible for developing ML models.
On a 5-point Likert scale from ``Not at all'' to ``Extremely'' useful, 52\% of the respondents who were asked the question (79\%) indicated that tools to facilitate 
communication between model developers and data collectors would be ``Very'' or ``Extremely'' useful.

Interviewees also shared a range of needs for tools
and processes
that can actively guide data collection
as it occurs,
to support fairness in downstream ML models. For example, in response to the ``oracle'' interview question, an ML engineer (R19) working on automated essay scoring noted:
\begin{quote}
\emph{``To score African American students fairly, they need examples of African American students scoring highly. But in the data [the data collection team] collect[s], this is very rare. So, what is the right way to sample [high-scorers] without having to score all the essays? [...] So [we need] some kind of way... to indicate [which schools] to collect from [...] or what to bother spending the extra money to score.''}
\end{quote}
A majority (60\%) of survey respondents, out of the 25\% who indicated their team has some control over data collection processes, indicated having such active guidance would be at least ``Very'' useful.
By contrast, in contexts where training data is collected via regular, ``in-the-wild'' users of a product, challenges can arise when specific user populations are less engaged with the product.
To overcome such challenges, a technical manager working on speech recognition (R31) 
suggested it would help to know more effective strategies to incentivize``in-the-wild''
product usage within specific populations, such as targeted gamification techniques~\cite{jia2016personality}.

\subsubsection{Scaffolding fairness-aware test set design}
Several interviewees stressed the central importance of careful test set design to detecting potential fairness issues. For example, R4's team would discuss possible issues to watch out for, and then
\emph{``try to design the test set to capture those notions, if we can.''}
R4 credited having
\emph{``the test set be well constructed and not biased''}
for the discovery of gender biases in their image captioner (e.g., images of female doctors were often mislabeled as nurses) which they ultimately traced to imbalances in their training data. An ML engineer working on gesture recognition (R6) noted it would be helpful to have better tools to scaffold the design of test sets, to make it easier to
\begin{quote}
\emph{``
assign tags to data points based on certain characteristics that you want to make sure are fair
[...and check]
that first of all, each of those tags has a significant number of samples
[and]
look at the measurements across each slice and [check] if there's a bias issue.''}
\end{quote}
A majority (66\%) of survey respondents, out of the 70\% who were asked this question, indicated that such tools to scaffold the design of test sets
would be ``Very''  or ``Extremely'' useful.

\subsubsection{Implications}
Although the fair ML literature has focused heavily on algorithmic ``de-biasing'' methods that assume a fixed training dataset, practitioners in many industry contexts have some control over data collection and curation.
Future research should support these practitioners in collecting and curating high quality datasets in the first place (cf.~\cite{chen2018my,kallus2018residual,gebru2018datasheets}). 
In contrast to the fair ML literature, HCI research on ML developer tools has often focused on the design of user interfaces to scaffold data collection and curation processes (e.g.,~\cite{chen2018anchorviz,chang2017revolt,gebru2017scalable,kulesza2014structured}). However, this work has tended to focus on improving ML models' overall predictive accuracy, rather than fairness.
A promising direction for future research
is to explore how such tools might be explicitly designed to support fairness and equity in downstream ML models by interactively guiding data collection, curation, and augmentation, or by supporting the design and use of test sets that can effectively surface biases and unfairness (cf.~\cite{bozonier2015test,janzen2005test,yang2018grounding}).\looseness=-1

%


\subsection{Challenges Due to Blind Spots} \label{sec:results-blindspots} 

Interviewees expressed many anxieties around their teams' potential ``blind spots,'' which might stand in the way of effectively addressing fairness issues, or even thinking to monitor for some forms of unfairness in the first place.

\subsubsection{Data collection and curation challenges due to blind spots}
Most of our interviewees highlighted needs for support in identifying which subpopulations their team needs to consider when developing specific kinds of ML applications, to ensure they collect sufficient data from these subpopulations or balance across them when curating existing datasets.
A technical director working on general-purpose ML tools (R32) emphasized the challenges of anticipating which subpopulations to consider, noting that this can be highly context and application dependent~\cite{greenhumyth}, with potential subpopulations extending well beyond those commonly discussed in the fair ML literature:
\emph{``Most of the time, people start thinking about attributes like
[ethnicity and gender...]. But the biggest problem I found is that these cohorts should be defined based on the domain and problem.  For example, for [automated writing evaluation] maybe it should be defined
based on [...whether the person is] a native speaker.''}
%
A majority (62\%) of survey respondents, out of the 80\% who were asked the question, indicated that it would be at least ``Very'' useful to have additional support in identifying relevant subpopulations. 

\subsubsection{"You'll know if there's fairness issues if someone raises hell online."\nopunct}

Interviewees often reported that their teams do not discover serious fairness issues until they receive customer complaints about products (or, worse, by reading negative media coverage about their products).
As a software engineer working on image classification (R7) put it, \emph{``How do you know the unknowns that you're being unfair towards? [...] You just have to put your model out there, and then you'll know if there's fairness issues if someone raises hell online.''}
Several interviewees expressed needs for support in detecting biases and unfairness prior to deployment, 
even in cases where they may not have anticipated 
all relevant subpopulations or all kinds of fairness issues. Despite their efforts running user studies, 
teams often discovered serious issues only after deploying a system in the real world (51\% of survey respondents marked this statement as at least ``Very'' accurate).

\subsubsection{Team biases and limitations}
Several of our interviewees' teams have a practice of getting together and trying to imagine everything that could go wrong with their products, so that they can make sure to proactively monitor for those issues. A few teams even reported including fairness-focused quizzes in their interview processes, with the aim of hiring employees who would be effective at spotting biases and unfairness.
R2 described much of their team's current process as
\emph{``just everyone collecting all the things that they can think of that could be offensive and testing for [them].''}
But as R4 emphasized,
\emph{``no one person on the team [has expertise] in all types of bias [...] especially when you take into account different cultures.''}
Interviewees noted that it would be helpful to somehow pool knowledge of potential fairness issues in specific application domains across teams with different backgrounds, who have complementary knowledge and blind spots.
A majority (67\%) of survey respondents, out of the 71\% who were asked the question, indicated that tools to support such knowledge pooling would be at least ``Very'' useful.\looseness=-1

A few interviewees also shared experiences in which efforts to obtain additional training data to address a fairness issue were hampered by their teams' 
blind spots. 
A developer working on image captioning (R4) recalled cases where many customers had complained that a globally deployed system performed well for celebrities from some countries, but routinely misidentified major celebrities from others:

\begin{quote}
\emph{``It sounds easy to just say like, `Oh, just add some more images in there,' but [...]
there's no person on the team that actually knows what all of [these celebrities] look like, for real [...] if I noticed that there's some celebrity from Taiwan that doesn't have enough images in there, I actually don't know what they look like to go and fix that.
[...]
But, \mbox{Beyonc{\'e}}, I know what she looks like.''}
\end{quote}

\subsubsection{Implications}
Assessing and mitigating unfairness in ML systems can depend on nuanced cultural and domain knowledge that no single product team is likely to have. In cases where ML products are deployed globally, efforts to recruit more diverse teams may be helpful yet insufficient.
A fruitful area for future research is the design of processes and tools to support effective sharing and re-use of such knowledge across team or company boundaries---for example via shared test sets (cf.~\cite{yang2018grounding}) or case studies from other teams who have worked in specific application domains. Another promising direction may be to support teams in the ad-hoc recruitment of diverse, team-external ``experts'' for particular tasks~\cite{valentine2017flash}.\looseness=-1

%
%


\subsection{Needs for More Proactive Auditing Processes} \label{sec:results-processes} 

Detecting potential fairness issues presents many unique auditing challenges. 
Interviewees often described their teams' current fairness auditing processes as reactive---with efforts tightly focused around specific customer complaints---in contrast to their teams' proactive approaches for detecting potential security risks. 
%
As a PM for 
web search (R11) put it,
\begin{quote}
\emph{``It's a little bit of a... manual search to say, `hey, we think this has a bias, let's go take a look and see if it does,' which I don't know is the right approach [...] because there are a lot of strange ones that you wouldn't expect [...] that we just accidentally stumbled upon.''}
\end{quote}
Out of the 63\% of survey respondents who were asked the question, almost half (49\%) reported that their team had previously found potential fairness issues in their products. Of the 51\%
who
indicated that
their team had not found any issues, most (80\%) suspected that there might be issues they had not yet detected, with a majority (55\%) reporting they believe undetected issues ``Probably'' or ``Definitely'' exist.

Furthermore, most interviewees noted that they are not generally rewarded within their organizations for their efforts around fairness. On a 5-point Likert scale from ``Not at all'' to ``A great deal,'' only 21\% of survey respondents reported that their team prioritizes fairness ``A lot'' or ``A great deal,'' and 36\% indicated ``Not at all.'' Given that interviewees often engaged in ML fairness efforts on their own time and initiative, several of the needs they highlighted focused on ways to reduce the amount of manual labor required to effectively monitor for unfairness, or ways to persuade other team members that a given fairness issue actually exists (e.g., by showing them quantitative metrics) and should be addressed.\looseness=-1

Interviewees revealed a range of needs for support in implementing proactive auditing processes.
These include
domain-specific auditing processes, including metrics and tools;
methods to effectively monitor for unfairness when individual-level demographics are unavailable;
more scalable and comprehensive auditing processes; and ways to determine if a specific issue is part of a systemic problem.\looseness=-1

\subsubsection{Needs for (domain-specific) standard auditing processes}
To progress beyond \emph{``having each team do some sort of ad hoc [testing]''} (R4),
several interviewees expressed desires for greater sharing of guidelines and processes. 
As R30 stressed,
\begin{quote}
\emph{``If you're developing [a model], there is a type of checklist that you go through for accuracy and so on. But there isn't anything like that [for fairness], or at least it hasn't been disseminated. What we need is a good way of incorporating
[fairness]
as part of the workflow.''}
\end{quote}

Most of our interviewees' teams do not currently have fairness metrics against which they can monitor performance and progress.
Only 23\% and 20\% of survey respondents, respectively, out of the 26\% who were asked these questions, marked as at least ``Very'' accurate the statements \emph{``We have [fairness] metrics / key performance indicators (KPIs)''} and \emph{``We run automated tests [for fairness].''}
Yet, as R2 noted,
\emph{``it's really hard to fix things that you can't measure.''}
Similarly, 
R1
said,
\emph{``it would be really nice to learn more about how unfair we actually are, because only then can we start tackling that.''}

Several of our interviewees reported that their team had searched the fair ML literature for existing fairness metrics. However, they often failed to find metrics that readily applied to their specific application domains. For example, although much of the fair ML literature has 
focused on metrics for quantifying ``allocative harms'' (relating to the allocation of limited opportunities, resources, or information), many of the fairness issues that arise in domains such as web search, conversational AI, or image captioning are ``representational harms'' (e.g., perpetuating undesirable associations)~\cite{crawford_2017}.\looseness=-1

Some interviewees reported that they had tried to hold meetings with other teams within their company, to learn from one another's experiences 
and avoid duplicated effort. However, as R2 explained, even when practitioners are working on different problems in the same application domain,
\begin{quote}
\emph{``It doesn't necessarily result in a best practices list. [...] We've all tried to make these ways to measure [unfairness ... but]  with each problem comes nuances that make it difficult to have one general way of testing.''}
\end{quote}
Given that most interviewees had limited time and resources to devote to developing their own solutions, they often emphasized the value of resources that could help them learn from others' experiences more efficiently.
For example, R21 suggested it would be ideal to have
\emph{``a nice white paper that's just like... `Here's a summary of research people have done on fairness [specifically] in NLP models.'\thinspace{}''}
Other interviewees suggested it would be extremely helpful to have access to tools and resources that would help their team anticipate what kinds of fairness issues can arise in their specific application domain,
together with domain-specific frameworks that can help them navigate any associated complexities.

\subsubsection{Fairness auditing without access to individual-level demographics}
Although most auditing methods in the fair ML literature assume access to sensitive demographics (such as gender or race) at an individual level~\cite{veale2017fairer}, many of our interviewees reported that their teams are only able to collect such information at coarser levels, if at all (cf.~\cite{veale2017fairer,kilbertus2018blind,barocas2016big}).
For example, companies working with K-12 student populations in the US are typically prohibited from collecting such demographics by school or district policies and FERPA laws~\cite{home_2018}.
A majority (70\%) of survey respondents, out of the 69\% who were asked the question, indicated that the availability of tools to support fairness auditing without access to demographics at an individual level would be at least ``Very'' useful.\looseness=-1

A few interviewees reported that their teams had attempted to use coarse-grained demographic information (e.g., region- or organization-level demographics) for fairness auditing.
However, each of these teams had quickly abandoned these efforts, citing limited time and resources to spend on building their own solutions.
R21 said,
\emph{``If we had more people who we could throw at this... `Can we leverage this fuzzy data to [audit]?' that would be great [...] It's a fairly intimidating research problem I think, for us.''}
Other interviewees noted that, while it would be helpful to have support in efficiently using coarse-grained demographic information for fairness auditing, several challenges would remain. For example,
as R14 noted,  
\emph{``even when you have those data [...] you may know a bunch about the demographics of a school, but then, you know, it turns out [our product] is only used by the gifted [or remedial] students, and you may not have means [to check].''}\looseness=-1

Some interviewees shared that their teams had experimented with developing ML models to infer individual-level demographics from available proxies, so that they could then use these inferred demographics to audit their ML products. However, interviewees worried that the use of proxies may in itself introduce undesirable biases, introducing a need to audit the auditing tool. A data scientist working on automated hiring 
(R23) recounted a time when their team had developed such a tool, but ultimately decided against using it:\looseness=-1
\begin{quote}
\emph{``We called it the SETHtimator, a sex and ethnicity estimator.
[...with] one dataset, we [only] had a list of people's names and their IP addresses. So we were able to sort of cross-reference their IP addresses with a name database, and from there use a [classifier] to list a probability that someone with that name in that region would have a certain gender or ethnicity. [...] It's buggy. If there was a tool [...to] do this automatically and with a trusted data source... that would be super useful.''}
\end{quote}

Interviewees often commented that it would be ideal to be able to
\emph{``get the demographic information in the first place''} (R15).
Recent work in the fair ML literature has proposed encryption 
mechanisms that ensure any collected individual-level demographics can only be used for 
auditing~\cite{kilbertus2018blind,veale2017fairer}. But interviewees emphasized that, while such technical solutions are an important prerequisite, half the battle would lie in convincing stakeholders and policymakers that these mechanisms are truly secure and that the benefits outweigh the risks of a potential data leak. 
Anticipating such challenges, a couple of interviewees
expressed interest in mechanisms that might allow local decision makers, such as healthcare professionals in hospitals, 
to use their ``on the ground'' knowledge of individual-level demographics to improve fairness in an ML system
without revealing these demographics.


\subsubsection{Needs for greater scalability and comprehensiveness}

Interviewees often complained about limitations of their current auditing processes, given the enormous space of potential fairness issues.
For example, a UX researcher working on chatbots (R18) highlighted the challenges of recruiting a sizable, diverse sample of user-study participants as one reason their team sometimes fails to detect fairness issues early on:
\emph{``because of just logistics... we get [8 or 10] participants at a time, and even though we recruit for a `diverse' group... I mean, we're not representing everybody.''}
Similarly, R4 described their team's current user-testing practices as
\emph{``more of a spot check,''}
noting
\emph{``what I would rather have is a more comprehensive... full bias scan.''}
Drawing parallels to existing, automated tools that scan natural language datasets for potentially sensitive terms,
R1 suggested having tools that
\emph{``at least flag things that seem potentially `unfair' would be helpful.''}
While several other interviewees generated similar ideas,
they also often pointed out that developing scalable, automated processes would be a highly challenging research problem, given that fairness can be so context dependent. R5 emphasized that domains like image captioning or web search are particularly challenging, because fairness can depend jointly on the system's output and the user-provided input (e.g., a query).\looseness=-1

\subsubsection{Diagnosing systemic problems from specific issues}
Interviewees also highlighted challenges in diagnosing whether specific issues 
(e.g., complaints from customers)
are symptomatic of broader, systemic problems or just ``one-offs.'' A product and data manager for a
translation system (R1) said
\begin{quote}
\emph{``If an oracle was able to tell me, `look, this is a severe problem and I can give you a hundred examples [of this problem],' [...] then it's much easier internally to get enough people to accept this and to solve it. So having a process which gives you more data points where you mess up [in this way] would be really helpful.''}
\end{quote}
A majority (62\%) of survey respondents, out of the 70\% who were asked this question, indicated that tools to help find other instances of an issue would be at least ``Very'' useful.\looseness=-1


\subsubsection{Implications}
Given that fairness can be highly context and application dependent~\cite{greenhumyth,lee2018understanding}, there is an urgent need for domain-specific educational resources, metrics, processes, and tools to help practitioners navigate the unique challenges that can arise in their specific application domains. Such resources might include, for example, accessible summaries of state-of-the-art research around fairness in machine translation, or case studies of challenges faced by teams working on particular kinds of recommender systems (e.g., \cite{cramerCaseStudy}).

Future research should also explore processes and tools to support fairness auditing with access to only coarse-grained demographic information (e.g., neighborhood- or school-level demographics). Recent work~\cite{kilbertus2018blind,veale2017fairer} has begun to explore the design of encryption 
mechanisms that ensure any collected individual-level demographics can only be used for auditing.
However, our interviewees noted that in certain sensitive contexts, stakeholders may be unwilling to reveal individual-level demographics, even with such mechanisms.\looseness=-1

Although interviewees highlighted needs for more scalable and comprehensive auditing processes, they also often expressed skepticism that fairness auditing could be fully automated in their domain due to challenges in quantifying fairness. Therefore, in addition to developing domain-specific metrics and tools, a direction for future research may be to explore and evaluate human-in-the-loop approaches to fairness auditing that combine the strengths of automated methods and human inspection (cf.~\cite{tan2018investigating,chen2018anchorviz,holstein2018student,veale2018fairness}), perhaps aided by tools for visualization and guided exploration~\cite{lakkaraju2017identifying,chen2018anchorviz}.

Finally, at a broader, organizational level, future research should explore how teams and companies can best be motivated to adopt fairness as a central priority.
Efforts to develop easily-trackable
fairness metrics in particular application domains may help towards this goal in strongly metric-driven companies. Relatedly, future tools for fairness auditing should be designed to support practitioners in both (1) determining whether specific issues are instances of broader, systemic problems and (2) effectively persuading other team members that there are issues that need to be addressed.


\subsection{Needs for More Holistic Auditing Methods} \label{sec:results-holistic} 

Much of the existing fair ML literature has focused on domains such as recidivism prediction, automated hiring, and face recognition, where fairness can be at least partially understood in terms of well-defined quantitative metrics (e.g., between-group parity in error rates or decisions \cite{chouldechova2018case,kleinberg2016inherent,buolamwini2018gender}). However, interviewees working on applications involving richer, complex interactions between the user and the system---such as chatbots, automated writing evaluation, adaptive tutoring and mentoring, and web search---brought up needs for more holistic, system-level auditing methods.\looseness=-1

\subsubsection{Fairness as a system-level property}
Many interviewees noted disconnects between the way they tend to think about fairness in their application domain and the discourse they have observed in both the popular press and academic literature. 
For example, a technical manager working on adaptive learning technologies (R30) noted that their team does not think about fairness in terms of monitoring individual ML models for unfairness, but instead evaluating the real-world impacts of ML systems, mentioning that
\emph{``
If we think about educational interventions as analogous to medical interventions or drug trials [...] we know and [expect] a particular intervention will have different effects on different subpopulations.''}

In a complex, multi-component ML system, there is not always a clean mapping between performance metrics for an individual ML model and the system's utility for users. System components may interact with one another~\cite{nushi2017human,dwork2018group} and with other aspects of a system's design in ways that can be difficult to predict absent an empirical study with actual users (cf.~\cite{friedman1996bias}). Furthermore, it may not be straightforward to even define fair system behavior without first understanding users' expectations and beliefs about the system (cf.~\cite{lee2018understanding,lee2017algorithmic}).
For example, a PM working on web search (R11) shared that their team had previously experimented with ways to address a fairness issue involving image search (a search for the term ``CEO'' resulted predominantly in images of white men). However, through user studies, the team learned that many users were uncomfortable with the idea of the company ``manipulating'' search results, viewing this behavior as unethical:\looseness=-1
\begin{quote}
\emph{``Users right now are seeing [image search] as `We show you [an objective] window into [...] society,' whereas we do have a strong argument [instead] for, `We should show you as many different types of images as possible,
[to] 
try to get something for everyone.'''}
\end{quote}

\subsubsection{Needs for simulation-based approaches for complex systems}
In applications involving sequences of complex interactions between the user and the system, fairness can be heavily context dependent. 
R17 suggested it would be valuable to have ways to prototype conversational agents more rapidly, including methods for simulating conversational trajectories (cf.~\cite{jain2018user,zhao2018socially})
\emph{``and then find[ing] ways to automate the identification of risky conversation patterns that emerge.''}
Similarly, a data scientist working on adaptive mentoring software (R10) suggested that because their product involves a long-term feedback loop, it would be ideal to run it on a population of ``simulated mentees'' (cf.~\cite{maclellan2016apprentice}) to see if certain forms of personalization might
be harmful with respect to equity.

\subsubsection{Implications}
In applications involving richer, complex interactions between the users and the system, assessing fairness issues via de-contextualized quantitative metrics for individual ML models may be insufficient (cf. \cite{selbst2018fairness}). Future research should explore new approaches to auditing and prototyping ML systems (cf.~\cite{dove2017ux}), perhaps aided by simulation tools that can help developers anticipate sensitive contexts or real-world impacts of using an ML system (cf. \cite{liu2018delayed,googleDesign}).


\subsection{Addressing Detected Issues} \label{sec:results-deciding} 

Interviewees revealed a range of challenges and needs around debugging and remediation of fairness issues once detected. These included, among others, needs for support in
identifying the cheapest, most effective strategies to address particular issues; 
methods to estimate how much additional data to collect for particular subpopulations; processes to anticipate
potential 
trade-offs between specific definitions of fairness and other desiderata for an ML system (not limited to predictive accuracy); and frameworks to help navigate complex ethical decisions 
(e.g., the fairness of fairness interventions).\looseness=-1
%





\subsubsection{Needs for support in strategy selection}
Interviewees reported that their teams often struggle to isolate the causes of unexpected fairness issues, especially when working with ML models that the team consider to be ``black boxes.''
As a result, it is often difficult for teams to decide where to focus their efforts---switching to a different model, augmenting the training data in some way, collecting more or different kinds of data, post-processing outputs, changing the objective function, or something else (cf.~\cite{chen2018my}). Of the 21\% of survey respondents whose teams had previously tried to address detected issues, a majority (54\%) indicated that it would be at least ``Very'' useful to have better support in comparing specific strategies for addressing particular fairness issues. Different costs may be associated with different strategies in different contexts. For example, a developer working on image captioning (R7) noted that collecting additional data is typically a \emph{``last resort''} option for their team, given data collection costs. In contrast, a PM working on image captioning on a different team (R2) 
cited data collection as their team's default starting place when trying to address a fairness issue. These interviewees and others also shared experiences where their teams had wasted significant time and resources pursuing various dead ends. As such, interviewees highlighted several needs for ``fair ML debugging'' processes and tools (cf.~\cite{chen2018my,galhotra2017fairness,google_whatif_2018}) to support their teams in identifying the cheapest, yet most promising strategies to address particular issues.
For example, R1 highlighted needs for support in \emph{``identify[ing] the component where we mess up''} in complex, multi-component ML systems~\cite{nushi2017human}, and in deciding whether to focus their mitigation efforts on training data or on models.
A majority (63\%) of survey respondents indicated that tools to aid in these decisions (cf.~\cite{chen2018my,google_whatif_2018}), would be at least ``Very'' useful.\looseness=-1

\subsubsection{Avoiding unexpected side effects of fairness interventions}
In addition to costs, 
interviewees often cited fears of unexpected side effects as a deterrent to addressing fairness issues. R4 shared prior experiences where, after making changes to datasets or models to improve fairness, their system changed in subtle, unexpected ways that harmed users' experiences:
\begin{quote}
\emph{``Even if your [quantitative metrics] come out better... at the end of the day, it's really just different from what you had before
[...] and [users]
notice that for their particular scenario, it's different in a negative way.''}
\end{quote}
A majority (71\%) of survey respondents indicated that it would be at least ``Very'' useful to have tools to help their team understand potential UX side effects caused by a particular strategy for addressing a fairness issue. To minimize the risk of side effects, several teams had a practice of implementing many local, ``band-aid,'' fixes rather than trying to address the root cause of an issue. In some cases (e.g., when the issue is a ``one-off''), such local fixes may be sufficient. However, interviewees also reported that these fixes, such as censoring specific system outputs or responses to certain user inputs, sometimes resulted in other forms of unfairness (e.g., harms to certain user populations caused by the censoring itself).

\subsubsection{How much more data would we need to collect?}
In cases where teams had considered addressing a fairness issue by collecting additional data, interviewees often shared needs for support in estimating the minimum number of additional data points per subpopulation that they would need to address the issue (66\% of survey respondents indicated that they would find such support at least ``Very'' useful).
Most of our interviewees' teams currently rely on developer intuition. 
For example, a PM working on image captioning (R2) said,\looseness=-1
\begin{quote}
\emph{``It's just hope and trial and error... [the developers have] experimented so much with these models [...] that they can say `generally, [we will need] this much data to make an impact on this type of model to change things this much.'\thinspace{}''}
\end{quote}
However, a developer on the same team (R4) noted that their initial
estimates are often wrong, which can be costly,
\begin{quote}
\emph{``especially when it takes two weeks to get an answer. [...] I always would just really want to know how much was enough.''}
\end{quote}

\subsubsection{Concerns about the fairness of fairness interventions}
Interviewees often expressed unease with the idea that their teams' technical choices can have major societal impacts. For example, a technical director working on general-purpose ML tools (R32) said,
\emph{``[ML] models' main assumption [is] that the past is similar to the future. [...] if I don't want to have the same future, am I in the position to define the future for society or not?''}
Another interviewee (R6) expressed doubts about the morality of targeting specific subpopulations for additional data collection, even if this data collection ultimately serves to improve fairness for those subpopulations (cf.~\cite{raghavan2018externalities}):\looseness=-1
\begin{quote}
\emph{``Targeting people based on certain aspects of their person... I don't know how we would go about doing that
[in the]
most morally and ethically and even vaguely responsible way.''}
\end{quote}
Several interviewees suggested it would be helpful to have access to domain-specific resources, such as ethical frameworks and case studies, to guide their teams' ongoing efforts around fairness (55\% of survey respondents indicated that having access to such resources would be at least ``Very'' useful).

\subsubsection{Changes to the broader system design}
The fair ML research literature has tended to focus heavily on the development of algorithmic methods to assess and mitigate biases in individual ML models. However, interviewees emphasized that many fairness issues that arise in real-world ML systems may be most effectively addressed through changes to the broader system design. For example, R3 recalled a case where their image captioner was systematically mislabeling images of female doctors as ``nurses,'' in accordance with historical stereotypes. The team resolved the issue by replacing the system outputs ``nurse'' and ``doctor'' with the more generic ``healthcare professional.''
Several other interviewees described ``fail-soft'' design strategies their team employs to try to ensure that the worst-case harm is minimized (cf.~\cite{liu2002makebelieve,d2008developing}). As a PM for web search (R5) put it,
\emph{``Sometimes, you start with what you know won't cause more harm, and [then] iterate.''} R14 noted that when their team designs actions (e.g., personalized messages) taken in response to particular model outputs, they try to imagine the impacts these actions might have in specific false-positive and false-negative scenarios. Indeed, 40\% of survey respondents reported that they had tried such fail-soft strategies to address fairness issues.

\subsubsection{Implications}
Future research should explore educational resources, processes, and tools to help teams identify the cheapest, most effective strategies to address particular fairness issues. This might include helping teams estimate the minimum amount of additional data required to address a fairness issue, or helping teams anticipate trade-offs between definitions of fairness and other desiderata for an ML system, such as user satisfaction (moving beyond the fair ML literature's focus on trade-offs between fairness and predictive accuracy).
In some cases, the best option available to a team may be to refrain from applying a particular fairness intervention (e.g., to avoid greater harms to users) \cite{selbst2018fairness}.


\subsection{Biases in the Humans in the Loop} \label{sec:results-biases} 

Finally, several interviewees stressed the importance of explicitly considering biases that may be present in the humans embedded in the various stages of the ML development pipeline, such as crowdworkers who annotate training data or user-study participants tasked with surfacing undesirable biases in ML systems \cite{attenberg2011beat,holstein2018classroom,kamar2016directions,nushi2017human,vaughan2017making}. For example, a UX designer working on automated essay scoring (R20) noted that their training data is collected by hiring human scorers to evaluate essays according to a detailed rubric. However, their team suspects that irrelevant factors may influence scorers' judgments \cite{amorim2018automated,malouff2016bias}. An ML engineer on the team (R19) suggested it would be valuable to have support in auditing their human scoring process.
R19 proposed auditing scorers by injecting artificially generated essays into the scoring pool, using a hypothetical tool that can
\emph{
``paraphrase [an essay] in another subgroup's style [...] a different voice [or] vernacular [...without] chang[ing] the linguistic content otherwise... and say, `If you apply this linguistic feature, do the scores change?'\thinspace{}''}
Similarly, 68\% of survey respondents marked tools to simulate counterfactuals (cf.~\cite{kusner2017counterfactual,google_whatif_2018}) as at least ``Very'' useful.\looseness=-1

More broadly, 69\% of survey respondents, out of the 79\% who were asked the question, marked tools to reduce the influence of human biases on their labeling or scoring processes (cf.~\cite{kamar2015identifying}) at least ``Very'' useful.
Furthermore, 69\% of respondents, out of the 32\% who were asked the question, reported that their teams already actively try to mitigate biases in their labeling or scoring processes at least ``Sometimes.''

\subsubsection{Implications}
Future research should explore ways to help teams better understand biases that may be present in the humans embedded throughout the ML development pipeline, as well as ways to mitigate these biases (see~\cite{kamar2015identifying,lyu2018sloppiness,vaughan2017making}).

%


\section{Conclusions and Future Directions} \label{sec:conclusion} 

Although industry practitioners are already grappling with biases and unfairness in ML systems, research on fair ML is rarely guided by an understanding of the challenges faced by practitioners~\cite{springerGarciaCramer2018,veale2018fairness}.
In this work, we conducted the first systematic investigation of industry teams' challenges and needs for support in developing fairer ML systems.
Even when practitioners are motivated to improve fairness in their products, they often face various technical and organizational barriers.
Below, we highlight just a few broad directions for future research to reduce some of these barriers:
\looseness=-1

\begin{itemize}[leftmargin=10pt]
\item Although the fair ML literature has overwhelmingly focused on algorithmic ``de-biasing,'' future research should also support practitioners in collecting and curating high-quality datasets in the first place, with an eye towards fairness in downstream ML models
(cf. ~\cite{chen2018my,kallus2018residual,gebru2018datasheets,liu2015shift,kamar2015identifying}).

\item Because fairness can be context and application dependent~\cite{greenhumyth,lee2018understanding}, domain-specific educational resources, metrics, processes, and tools (e.g., \cite{Bender2018data,cramerCaseStudy}) are urgently needed.\looseness=-1

\item Although most fairness auditing methods assume access to individual-level demographics, many teams are only able to collect such information at coarser levels, if at all. Future research should explore ways to support fairness auditing with access to only coarse-grained demographic information (e.g., neighborhood- or school-level demographics).

\item Another rich area for future research is the development of processes and tools for fairness-focused debugging~\cite{galhotra2017fairness,google_whatif_2018}. For instance, it can be highly challenging to determine whether specific issues (e.g., complaints from customers) are ``one-offs'' or symptomatic of broader, systemic problems that might require deeper investigation, let alone to identify the most effective strategies for addressing them.\looseness=-1

\item Finally, our findings point to needs for automated auditing tools and new approaches to prototyping ML systems (cf.~\cite{dove2017ux}). 
Interviewees highlighted limitations of existing UX prototyping methods for surfacing fairness issues in complex, multi-component ML systems (e.g., chatbots), where fairness can be heavily context dependent~\cite{greenhumyth,lee2018understanding}, and the space of potential contexts is often very large.
%

\end{itemize}

The rapidly growing area of fairness in ML presents many new challenges. ML systems are increasingly widespread, 
with demonstrated potential to amplify social inequities, or even to create new ones~\cite{angwin2016machine,bbc_news_2013,bolukbasi2016man,buolamwini2018gender,reisman2018algorithmic}.
Therefore, as research in this area progresses, it is urgent that research agendas be aligned with the challenges and 
needs of those who affect and are affected by ML systems.
We view the directions outlined in this paper as critical opportunities for the ML and HCI research communities to play more active, collaborative roles in mitigating unfairness in real-world ML systems.\looseness=-1


\begin{acks}
We thank all our interviewees and survey respondents for their participation. We also thank Mary Beth Kery, Bogdan Kulynych, Michael Madaio, Alexandra Olteanu, Rebekah Overdorf, Ronni Sadovsky, Joseph Seering, Ben Shneiderman, Michael Veale, and our anonymous CHI 2019 reviewers for their insightful feedback. This research was initiated while the first author was a summer intern at Microsoft Research.
\end{acks}

\balance

\bibliographystyle{ACM-Reference-Format}
\bibliography{sample-bibliography}


\begin{thebibliography}{113}


\ifx \showCODEN    \undefined \def \showCODEN     #1{\unskip}     \fi
\ifx \showDOI      \undefined \def \showDOI       #1{#1}\fi
\ifx \showISBNx    \undefined \def \showISBNx     #1{\unskip}     \fi
\ifx \showISBNxiii \undefined \def \showISBNxiii  #1{\unskip}     \fi
\ifx \showISSN     \undefined \def \showISSN      #1{\unskip}     \fi
\ifx \showLCCN     \undefined \def \showLCCN      #1{\unskip}     \fi
\ifx \shownote     \undefined \def \shownote      #1{#1}          \fi
\ifx \showarticletitle \undefined \def \showarticletitle #1{#1}   \fi
\ifx \showURL      \undefined \def \showURL       {\relax}        \fi
\providecommand\bibfield[2]{#2}
\providecommand\bibinfo[2]{#2}
\providecommand\natexlab[1]{#1}
\providecommand\showeprint[2][]{arXiv:#2}

\bibitem[\protect\citeauthoryear{ACM}{ACM}{2018a}]%
        {FAT*}
\bibfield{author}{\bibinfo{person}{ACM}.} \bibinfo{year}{2018}\natexlab{a}.
\newblock \bibinfo{title}{ACM Conference on Fairness, Accountability, and
  Transparency (ACM FAT*)}.
\newblock
\newblock
\urldef\tempurl%
\url{https://fatconference.org/}
\showURL{%
\tempurl}
\newblock
\shownote{Accessed: 2018-06-15.}


\bibitem[\protect\citeauthoryear{ACM}{ACM}{2018b}]%
        {FATML}
\bibfield{author}{\bibinfo{person}{ACM}.} \bibinfo{year}{2018}\natexlab{b}.
\newblock \bibinfo{title}{FAT/ML}.
\newblock
\newblock
\newblock
\shownote{https://www.fatml.org.Accessed: 2018-06-15.}


\bibitem[\protect\citeauthoryear{Agarwal, Beygelzimer, Dud{\'\i}k, Langford,
  and Wallach}{Agarwal et~al\mbox{.}}{2018}]%
        {agarwal2018reductions}
\bibfield{author}{\bibinfo{person}{Alekh Agarwal}, \bibinfo{person}{Alina
  Beygelzimer}, \bibinfo{person}{Miroslav Dud{\'\i}k}, \bibinfo{person}{John
  Langford}, {and} \bibinfo{person}{Hanna Wallach}.}
  \bibinfo{year}{2018}\natexlab{}.
\newblock \showarticletitle{A reductions approach to fair classification}. In
  \bibinfo{booktitle}{\emph{Proceedings of the Thirty-fifth International
  Conference on Machine Learning (ICML 2018)}}.
\newblock


\bibitem[\protect\citeauthoryear{Alvarado and Waern}{Alvarado and
  Waern}{2018}]%
        {alvarado2018towards}
\bibfield{author}{\bibinfo{person}{Oscar Alvarado} {and}
  \bibinfo{person}{Annika Waern}.} \bibinfo{year}{2018}\natexlab{}.
\newblock \showarticletitle{Towards algorithmic experience: Initial efforts for
  social media contexts}. In \bibinfo{booktitle}{\emph{Proceedings of the 2018
  CHI Conference on Human Factors in Computing Systems (CHI 2018)}}. ACM,
  \bibinfo{pages}{286}.
\newblock


\bibitem[\protect\citeauthoryear{Amershi, Chickering, Drucker, Lee, Simard, and
  Suh}{Amershi et~al\mbox{.}}{2015}]%
        {amershi2015modeltracker}
\bibfield{author}{\bibinfo{person}{Saleema Amershi}, \bibinfo{person}{Max
  Chickering}, \bibinfo{person}{Steven~M Drucker}, \bibinfo{person}{Bongshin
  Lee}, \bibinfo{person}{Patrice Simard}, {and} \bibinfo{person}{Jina Suh}.}
  \bibinfo{year}{2015}\natexlab{}.
\newblock \showarticletitle{ModelTracker: Redesigning performance analysis
  tools for machine learning}. In \bibinfo{booktitle}{\emph{Proceedings of the
  2015 CHI Conference on Human Factors in Computing Systems (CHI 2015)}}. ACM,
  \bibinfo{pages}{337--346}.
\newblock


\bibitem[\protect\citeauthoryear{Amorim, Can{\c{c}}ado, and Veloso}{Amorim
  et~al\mbox{.}}{2018}]%
        {amorim2018automated}
\bibfield{author}{\bibinfo{person}{Evelin Amorim}, \bibinfo{person}{Marcia
  Can{\c{c}}ado}, {and} \bibinfo{person}{Adriano Veloso}.}
  \bibinfo{year}{2018}\natexlab{}.
\newblock \showarticletitle{Automated essay scoring in the presence of biased
  ratings}. In \bibinfo{booktitle}{\emph{Proceedings of the 2018 Conference of
  the North American Chapter of the Association for Computational Linguistics:
  Human Language Technologies, Volume 1 (Long Papers)}},
  Vol.~\bibinfo{volume}{1}. \bibinfo{pages}{229--237}.
\newblock


\bibitem[\protect\citeauthoryear{Angell, Johnson, Brun, and Meliou}{Angell
  et~al\mbox{.}}{2018}]%
        {angell2018themis}
\bibfield{author}{\bibinfo{person}{Rico Angell}, \bibinfo{person}{Brittany
  Johnson}, \bibinfo{person}{Yuriy Brun}, {and} \bibinfo{person}{Alexandra
  Meliou}.} \bibinfo{year}{2018}\natexlab{}.
\newblock \showarticletitle{Themis: Automatically testing software for
  discrimination}. In \bibinfo{booktitle}{\emph{Proceedings of the
  Demonstrations Track at the 26th ACM Joint European Software Engineering
  Conference and Symposium on the Foundations of Software Engineering
  (ESEC/FSE), Lake Buena Vista, FL, USA}}.
\newblock


\bibitem[\protect\citeauthoryear{Angwin, Larson, Mattu, and Kirchner}{Angwin
  et~al\mbox{.}}{2016}]%
        {angwin2016machine}
\bibfield{author}{\bibinfo{person}{Julia Angwin}, \bibinfo{person}{Jeff
  Larson}, \bibinfo{person}{Surya Mattu}, {and} \bibinfo{person}{Lauren
  Kirchner}.} \bibinfo{year}{2016}\natexlab{}.
\newblock \showarticletitle{Machine bias: There's software used across the
  country to predict future criminals, and it's biased against blacks}.
\newblock \bibinfo{journal}{\emph{ProPublica}} (\bibinfo{year}{2016}).
\newblock


\bibitem[\protect\citeauthoryear{Attenberg, Ipeirotis, and Provost}{Attenberg
  et~al\mbox{.}}{2011}]%
        {attenberg2011beat}
\bibfield{author}{\bibinfo{person}{Josh Attenberg},
  \bibinfo{person}{Panagiotis~G Ipeirotis}, {and} \bibinfo{person}{Foster~J
  Provost}.} \bibinfo{year}{2011}\natexlab{}.
\newblock \showarticletitle{Beat the machine: Challenging workers to find the
  unknown unknowns}.
\newblock \bibinfo{journal}{\emph{Human Computation}} \bibinfo{volume}{11},
  \bibinfo{number}{11} (\bibinfo{year}{2011}), \bibinfo{pages}{2--7}.
\newblock


\bibitem[\protect\citeauthoryear{Baker, Corbett, Roll, and Koedinger}{Baker
  et~al\mbox{.}}{2008}]%
        {d2008developing}
\bibfield{author}{\bibinfo{person}{Ryan~SJd Baker}, \bibinfo{person}{Albert~T
  Corbett}, \bibinfo{person}{Ido Roll}, {and} \bibinfo{person}{Kenneth~R
  Koedinger}.} \bibinfo{year}{2008}\natexlab{}.
\newblock \showarticletitle{Developing a generalizable detector of when
  students game the system}.
\newblock \bibinfo{journal}{\emph{User Modeling and User-Adapted Interaction}}
  \bibinfo{volume}{18}, \bibinfo{number}{3} (\bibinfo{year}{2008}),
  \bibinfo{pages}{287--314}.
\newblock


\bibitem[\protect\citeauthoryear{Barocas and Selbst}{Barocas and
  Selbst}{2016}]%
        {barocas2016big}
\bibfield{author}{\bibinfo{person}{Solon Barocas} {and}
  \bibinfo{person}{Andrew~D Selbst}.} \bibinfo{year}{2016}\natexlab{}.
\newblock \showarticletitle{Big data's disparate impact}.
\newblock \bibinfo{journal}{\emph{Cal. L. Rev.}}  \bibinfo{volume}{104}
  (\bibinfo{year}{2016}), \bibinfo{pages}{671}.
\newblock


\bibitem[\protect\citeauthoryear{BBC}{BBC}{2013}]%
        {bbc_news_2013}
\bibfield{author}{\bibinfo{person}{BBC}.} \bibinfo{year}{2013}\natexlab{}.
\newblock \showarticletitle{Google searches expose racial bias, says study of
  names}.
\newblock \bibinfo{journal}{\emph{BBC News}} (\bibinfo{date}{Feb}
  \bibinfo{year}{2013}).
\newblock
\newblock
\shownote{https://www.bbc.com/news/technology-21322183. Accessed: 2018-09-03.}


\bibitem[\protect\citeauthoryear{Bender and Friedman}{Bender and
  Friedman}{2018}]%
        {Bender2018data}
\bibfield{author}{\bibinfo{person}{Emily Bender} {and} \bibinfo{person}{Batya
  Friedman}.} \bibinfo{year}{2018}\natexlab{}.
\newblock \showarticletitle{Data statements for NLP: Toward mitigating system
  bias and enabling better science}.
\newblock \bibinfo{journal}{\emph{OpenReview Preprint}}.
\newblock


\bibitem[\protect\citeauthoryear{Berk, Heidari, Jabbari, Kearns, and Roth}{Berk
  et~al\mbox{.}}{2018}]%
        {berk2017fairness}
\bibfield{author}{\bibinfo{person}{Richard Berk}, \bibinfo{person}{Hoda
  Heidari}, \bibinfo{person}{Shahin Jabbari}, \bibinfo{person}{Michael Kearns},
  {and} \bibinfo{person}{Aaron Roth}.} \bibinfo{year}{2018}\natexlab{}.
\newblock \showarticletitle{Fairness in criminal justice risk assessments: The
  State of the Art}.
\newblock \bibinfo{journal}{\emph{Sociological Methods \& Research}}
  (\bibinfo{year}{2018}).
\newblock


\bibitem[\protect\citeauthoryear{Binns}{Binns}{2018}]%
        {binns2017fairnesspoliticalphilosophy}
\bibfield{author}{\bibinfo{person}{Reuben Binns}.}
  \bibinfo{year}{2018}\natexlab{}.
\newblock \showarticletitle{Fairness in machine learning: Lessons from
  political philosophy}.
\newblock \bibinfo{journal}{\emph{Proceedings of Machine Learning Research}}
  \bibinfo{volume}{81}, \bibinfo{pages}{149--159}.
\newblock


\bibitem[\protect\citeauthoryear{Binns, Van~Kleek, Veale, Lyngs, Zhao, and
  Shadbolt}{Binns et~al\mbox{.}}{2018}]%
        {binns2018s}
\bibfield{author}{\bibinfo{person}{Reuben Binns}, \bibinfo{person}{Max
  Van~Kleek}, \bibinfo{person}{Michael Veale}, \bibinfo{person}{Ulrik Lyngs},
  \bibinfo{person}{Jun Zhao}, {and} \bibinfo{person}{Nigel Shadbolt}.}
  \bibinfo{year}{2018}\natexlab{}.
\newblock \showarticletitle{``It's reducing a human being to a percentage'':
  Perceptions of justice in algorithmic decisions}. In
  \bibinfo{booktitle}{\emph{Proceedings of the 2018 CHI Conference on Human
  Factors in Computing Systems (CHI 2018)}}. ACM, \bibinfo{pages}{377}.
\newblock


\bibitem[\protect\citeauthoryear{Bolukbasi, Chang, Zou, Saligrama, and
  Kalai}{Bolukbasi et~al\mbox{.}}{2016}]%
        {bolukbasi2016man}
\bibfield{author}{\bibinfo{person}{Tolga Bolukbasi}, \bibinfo{person}{Kai-Wei
  Chang}, \bibinfo{person}{James~Y Zou}, \bibinfo{person}{Venkatesh Saligrama},
  {and} \bibinfo{person}{Adam~T Kalai}.} \bibinfo{year}{2016}\natexlab{}.
\newblock \showarticletitle{Man is to computer programmer as woman is to
  homemaker? Debiasing word embeddings}. In \bibinfo{booktitle}{\emph{Advances
  in Neural Information Processing Systems (NeurIPS 2016)}}.
  \bibinfo{pages}{4349--4357}.
\newblock


\bibitem[\protect\citeauthoryear{Bosch, D'Mello, Baker, Ocumpaugh, Shute,
  Ventura, Wang, and Zhao}{Bosch et~al\mbox{.}}{2016}]%
        {bosch2016detecting}
\bibfield{author}{\bibinfo{person}{Nigel Bosch}, \bibinfo{person}{Sidney~K
  D'Mello}, \bibinfo{person}{Ryan~S Baker}, \bibinfo{person}{Jaclyn Ocumpaugh},
  \bibinfo{person}{Valerie Shute}, \bibinfo{person}{Matthew Ventura},
  \bibinfo{person}{Lubin Wang}, {and} \bibinfo{person}{Weinan Zhao}.}
  \bibinfo{year}{2016}\natexlab{}.
\newblock \showarticletitle{Detecting student emotions in computer-enabled
  classrooms}. In \bibinfo{booktitle}{\emph{Proceedings of the 2016
  International Joint Conference on Artificial Intelligence (IJCAI 2016)}}.
  \bibinfo{pages}{4125--4129}.
\newblock


\bibitem[\protect\citeauthoryear{Bozonier}{Bozonier}{2015}]%
        {bozonier2015test}
\bibfield{author}{\bibinfo{person}{Justin Bozonier}.}
  \bibinfo{year}{2015}\natexlab{}.
\newblock \bibinfo{booktitle}{\emph{Test-driven machine learning}}.
\newblock \bibinfo{publisher}{Packt Publishing Ltd}.
\newblock


\bibitem[\protect\citeauthoryear{Bucher}{Bucher}{2017}]%
        {bucher2017algorithmic}
\bibfield{author}{\bibinfo{person}{Taina Bucher}.}
  \bibinfo{year}{2017}\natexlab{}.
\newblock \showarticletitle{The algorithmic imaginary: Exploring the ordinary
  affects of Facebook algorithms}.
\newblock \bibinfo{journal}{\emph{Information, Communication \& Society}}
  \bibinfo{volume}{20}, \bibinfo{number}{1} (\bibinfo{year}{2017}),
  \bibinfo{pages}{30--44}.
\newblock


\bibitem[\protect\citeauthoryear{Buolamwini and Gebru}{Buolamwini and
  Gebru}{2018}]%
        {buolamwini2018gender}
\bibfield{author}{\bibinfo{person}{Joy Buolamwini} {and}
  \bibinfo{person}{Timnit Gebru}.} \bibinfo{year}{2018}\natexlab{}.
\newblock \showarticletitle{Gender Shades: Intersectional accuracy disparities
  in commercial gender classification}. In
  \bibinfo{booktitle}{\emph{Proceedings of the 2018 Conference on Fairness,
  Accountability and Transparency (FAT* 2018)}}. \bibinfo{pages}{77--91}.
\newblock


\bibitem[\protect\citeauthoryear{Chang, Amershi, and Kamar}{Chang
  et~al\mbox{.}}{2017}]%
        {chang2017revolt}
\bibfield{author}{\bibinfo{person}{Joseph~Chee Chang}, \bibinfo{person}{Saleema
  Amershi}, {and} \bibinfo{person}{Ece Kamar}.}
  \bibinfo{year}{2017}\natexlab{}.
\newblock \showarticletitle{Revolt: Collaborative crowdsourcing for labeling
  machine learning datasets}. In \bibinfo{booktitle}{\emph{Proceedings of the
  2017 CHI Conference on Human Factors in Computing Systems (CHI 2017)}}. ACM,
  \bibinfo{pages}{2334--2346}.
\newblock


\bibitem[\protect\citeauthoryear{Chen, Johansson, and Sontag}{Chen
  et~al\mbox{.}}{2018a}]%
        {chen2018my}
\bibfield{author}{\bibinfo{person}{Irene Chen}, \bibinfo{person}{Fredrik~D
  Johansson}, {and} \bibinfo{person}{David Sontag}.}
  \bibinfo{year}{2018}\natexlab{a}.
\newblock \showarticletitle{Why is my classifier discriminatory?}
\newblock \bibinfo{journal}{\emph{Advances in Neural Information Processing
  Systems (NeurIPS 2018)}}.
\newblock


\bibitem[\protect\citeauthoryear{Chen, Suh, Verwey, Ramos, Drucker, and
  Simard}{Chen et~al\mbox{.}}{2018b}]%
        {chen2018anchorviz}
\bibfield{author}{\bibinfo{person}{Nan-Chen Chen}, \bibinfo{person}{Jina Suh},
  \bibinfo{person}{Johan Verwey}, \bibinfo{person}{Gonzalo Ramos},
  \bibinfo{person}{Steven Drucker}, {and} \bibinfo{person}{Patrice Simard}.}
  \bibinfo{year}{2018}\natexlab{b}.
\newblock \showarticletitle{AnchorViz: Facilitating classifier error discovery
  through interactive semantic data exploration}. In
  \bibinfo{booktitle}{\emph{23rd International Conference on Intelligent User
  Interfaces (IUI 2018)}}. ACM, \bibinfo{pages}{269--280}.
\newblock


\bibitem[\protect\citeauthoryear{Chouldechova}{Chouldechova}{2017}]%
        {chouldechova2017fair}
\bibfield{author}{\bibinfo{person}{Alexandra Chouldechova}.}
  \bibinfo{year}{2017}\natexlab{}.
\newblock \showarticletitle{Fair prediction with disparate impact: A study of
  bias in recidivism prediction instruments}.
\newblock \bibinfo{journal}{\emph{Big Data}} \bibinfo{volume}{5},
  \bibinfo{number}{2} (\bibinfo{year}{2017}), \bibinfo{pages}{153--163}.
\newblock


\bibitem[\protect\citeauthoryear{Chouldechova, Benavides-Prado, Fialko, and
  Vaithianathan}{Chouldechova et~al\mbox{.}}{2018}]%
        {chouldechova2018case}
\bibfield{author}{\bibinfo{person}{Alexandra Chouldechova},
  \bibinfo{person}{Diana Benavides-Prado}, \bibinfo{person}{Oleksandr Fialko},
  {and} \bibinfo{person}{Rhema Vaithianathan}.}
  \bibinfo{year}{2018}\natexlab{}.
\newblock \showarticletitle{A case study of algorithm-assisted decision making
  in child maltreatment hotline screening decisions}. In
  \bibinfo{booktitle}{\emph{ACM Conference on Fairness, Accountability, and
  Transparency (FAT* 2018)}}. \bibinfo{pages}{134--148}.
\newblock


\bibitem[\protect\citeauthoryear{Cramer, Garcia-Gathright, Reddy, Springer, and
  Takeo}{Cramer et~al\mbox{.}}{ress}]%
        {cramerCaseStudy}
\bibfield{author}{\bibinfo{person}{Henriette Cramer}, \bibinfo{person}{Jean
  Garcia-Gathright}, \bibinfo{person}{Sravana Reddy}, \bibinfo{person}{Aaron
  Springer}, {and} \bibinfo{person}{Romain Takeo}.} \bibinfo{year}{In
  press}\natexlab{}.
\newblock \showarticletitle{Translation, tracks and data: Algorithmic bias in
  practice}.
\newblock \bibinfo{journal}{\emph{Extended Abstracts of the 2019 CHI Conference
  on Human Factors in Computing Systems (CHI EA 2019)}}.
\newblock


\bibitem[\protect\citeauthoryear{Crawford}{Crawford}{2017}]%
        {crawford_2017}
\bibfield{author}{\bibinfo{person}{Kate Crawford}.}
  \bibinfo{year}{2017}\natexlab{}.
\newblock \bibinfo{title}{Artificial intelligence with very real biases}.
\newblock
\newblock
\newblock
\shownote{http://www.wsj.com/articles/artificial-intelligencewith-very-real-biases-1508252717.
  Accessed: 2018-06-15.}


\bibitem[\protect\citeauthoryear{D{\'\i}az, Johnson, Lazar, Piper, and
  Gergle}{D{\'\i}az et~al\mbox{.}}{2018}]%
        {diaz2018addressing}
\bibfield{author}{\bibinfo{person}{Mark D{\'\i}az}, \bibinfo{person}{Isaac
  Johnson}, \bibinfo{person}{Amanda Lazar}, \bibinfo{person}{Anne~Marie Piper},
  {and} \bibinfo{person}{Darren Gergle}.} \bibinfo{year}{2018}\natexlab{}.
\newblock \showarticletitle{Addressing age-related bias in sentiment analysis}.
  In \bibinfo{booktitle}{\emph{Proceedings of the 2018 CHI Conference on Human
  Factors in Computing Systems (CHI 2018)}}. ACM, \bibinfo{pages}{412}.
\newblock


\bibitem[\protect\citeauthoryear{Dove, Halskov, Forlizzi, and Zimmerman}{Dove
  et~al\mbox{.}}{2017}]%
        {dove2017ux}
\bibfield{author}{\bibinfo{person}{Graham Dove}, \bibinfo{person}{Kim Halskov},
  \bibinfo{person}{Jodi Forlizzi}, {and} \bibinfo{person}{John Zimmerman}.}
  \bibinfo{year}{2017}\natexlab{}.
\newblock \showarticletitle{UX design innovation: Challenges for working with
  machine learning as a design material}. In
  \bibinfo{booktitle}{\emph{Proceedings of the 2017 CHI Conference on Human
  Factors in Computing Systems (CHI 2017)}}. ACM, \bibinfo{pages}{278--288}.
\newblock


\bibitem[\protect\citeauthoryear{DSSG}{DSSG}{2018}]%
        {aequitas}
\bibfield{author}{\bibinfo{person}{DSSG}.} \bibinfo{year}{2018}\natexlab{}.
\newblock \bibinfo{title}{Aequitas: Bias and fairness audit toolkit}.
\newblock
\newblock
\newblock
\shownote{http://aequitas.dssg.io. Accessed: 2018-08-29.}


\bibitem[\protect\citeauthoryear{Dwork, Hardt, Pitassi, Reingold, and
  Zemel}{Dwork et~al\mbox{.}}{2012}]%
        {dwork2012fairness}
\bibfield{author}{\bibinfo{person}{Cynthia Dwork}, \bibinfo{person}{Moritz
  Hardt}, \bibinfo{person}{Toniann Pitassi}, \bibinfo{person}{Omer Reingold},
  {and} \bibinfo{person}{Richard Zemel}.} \bibinfo{year}{2012}\natexlab{}.
\newblock \showarticletitle{Fairness through awareness}. In
  \bibinfo{booktitle}{\emph{Proceedings of the Third Innovations in Theoretical
  Computer Science Conference (ITCS 2012)}}. ACM, \bibinfo{pages}{214--226}.
\newblock


\bibitem[\protect\citeauthoryear{Dwork and Ilvento}{Dwork and Ilvento}{2018}]%
        {dwork2018group}
\bibfield{author}{\bibinfo{person}{Cynthia Dwork} {and}
  \bibinfo{person}{Christina Ilvento}.} \bibinfo{year}{2018}\natexlab{}.
\newblock \bibinfo{title}{Fairness under composition}.
\newblock
\newblock
\newblock
\shownote{CoRR arXiv:1806.06122.}


\bibitem[\protect\citeauthoryear{Esteva, Kuprel, Novoa, Ko, Swetter, Blau, and
  Thrun}{Esteva et~al\mbox{.}}{2017}]%
        {esteva2017dermatologist}
\bibfield{author}{\bibinfo{person}{Andre Esteva}, \bibinfo{person}{Brett
  Kuprel}, \bibinfo{person}{Roberto~A Novoa}, \bibinfo{person}{Justin Ko},
  \bibinfo{person}{Susan~M Swetter}, \bibinfo{person}{Helen~M Blau}, {and}
  \bibinfo{person}{Sebastian Thrun}.} \bibinfo{year}{2017}\natexlab{}.
\newblock \showarticletitle{Dermatologist-level classification of skin cancer
  with deep neural networks}.
\newblock \bibinfo{journal}{\emph{Nature}} \bibinfo{volume}{542},
  \bibinfo{number}{7639} (\bibinfo{year}{2017}), \bibinfo{pages}{115}.
\newblock


\bibitem[\protect\citeauthoryear{Ferryman and Pitcan}{Ferryman and
  Pitcan}{2018}]%
        {ferryman2018fairness}
\bibfield{author}{\bibinfo{person}{Kadija Ferryman} {and}
  \bibinfo{person}{Mikaela Pitcan}.} \bibinfo{year}{2018}\natexlab{}.
\newblock \showarticletitle{Fairness in precision medicine}.
\newblock \bibinfo{journal}{\emph{Data \& Society}} (\bibinfo{year}{2018}).
\newblock


\bibitem[\protect\citeauthoryear{Foundation}{Foundation}{2017}]%
        {WWWfoundation}
\bibfield{author}{\bibinfo{person}{World Wide~Web Foundation}.}
  \bibinfo{year}{2017}\natexlab{}.
\newblock \showarticletitle{Algorithmic accountability}.
\newblock \bibinfo{journal}{\emph{World Wide Web Foundation}}
  (\bibinfo{year}{2017}).
\newblock


\bibitem[\protect\citeauthoryear{Friedman and Nissenbaum}{Friedman and
  Nissenbaum}{1996}]%
        {friedman1996bias}
\bibfield{author}{\bibinfo{person}{Batya Friedman} {and} \bibinfo{person}{Helen
  Nissenbaum}.} \bibinfo{year}{1996}\natexlab{}.
\newblock \showarticletitle{Bias in computer systems}.
\newblock \bibinfo{journal}{\emph{ACM Transactions on Information Systems
  (TOIS)}} \bibinfo{volume}{14}, \bibinfo{number}{3} (\bibinfo{year}{1996}),
  \bibinfo{pages}{330--347}.
\newblock


\bibitem[\protect\citeauthoryear{Galhotra, Brun, and Meliou}{Galhotra
  et~al\mbox{.}}{2017}]%
        {galhotra2017fairness}
\bibfield{author}{\bibinfo{person}{Sainyam Galhotra}, \bibinfo{person}{Yuriy
  Brun}, {and} \bibinfo{person}{Alexandra Meliou}.}
  \bibinfo{year}{2017}\natexlab{}.
\newblock \showarticletitle{Fairness testing: Testing software for
  discrimination}. In \bibinfo{booktitle}{\emph{Proceedings of the 2017 11th
  Joint Meeting on Foundations of Software Engineering (FSE 2017)}}. ACM,
  \bibinfo{pages}{498--510}.
\newblock


\bibitem[\protect\citeauthoryear{Gebru, Krause, Deng, and Fei-Fei}{Gebru
  et~al\mbox{.}}{2017}]%
        {gebru2017scalable}
\bibfield{author}{\bibinfo{person}{Timnit Gebru}, \bibinfo{person}{Jonathan
  Krause}, \bibinfo{person}{Jia Deng}, {and} \bibinfo{person}{Li Fei-Fei}.}
  \bibinfo{year}{2017}\natexlab{}.
\newblock \showarticletitle{Scalable annotation of fine-grained categories
  without experts}. In \bibinfo{booktitle}{\emph{Proceedings of the 2017 CHI
  Conference on Human Factors in Computing Systems (CHI 2017)}}. ACM,
  \bibinfo{pages}{1877--1881}.
\newblock


\bibitem[\protect\citeauthoryear{Gebru, Morgenstern, Vecchione, Vaughan,
  Wallach, Daum{\'e}~III, and Crawford}{Gebru et~al\mbox{.}}{2018}]%
        {gebru2018datasheets}
\bibfield{author}{\bibinfo{person}{Timnit Gebru}, \bibinfo{person}{Jamie
  Morgenstern}, \bibinfo{person}{Briana Vecchione},
  \bibinfo{person}{Jennifer~Wortman Vaughan}, \bibinfo{person}{Hanna Wallach},
  \bibinfo{person}{Hal Daum{\'e}~III}, {and} \bibinfo{person}{Kate Crawford}.}
  \bibinfo{year}{2018}\natexlab{}.
\newblock \bibinfo{title}{Datasheets for datasets}.
\newblock
\newblock
\newblock
\shownote{CoRR arXiv:1803.09010.}


\bibitem[\protect\citeauthoryear{Gershgorn}{Gershgorn}{2018a}]%
        {gershgorn_2018c}
\bibfield{author}{\bibinfo{person}{Dave Gershgorn}.}
  \bibinfo{year}{2018}\natexlab{a}.
\newblock \bibinfo{title}{America's biggest body-camera company says facial
  recognition isn't accurate enough for police}.
\newblock
\newblock
\newblock
\shownote{https://qz.com/1351519/facial-recognition-isnt-yet-accurate-enough-for-policing-decisions/.
  Accessed: 2018-08-30.}


\bibitem[\protect\citeauthoryear{Gershgorn}{Gershgorn}{2018b}]%
        {gershgorn_2018b}
\bibfield{author}{\bibinfo{person}{Dave Gershgorn}.}
  \bibinfo{year}{2018}\natexlab{b}.
\newblock \bibinfo{title}{If AI is going to be the world's doctor, it needs
  better textbooks}.
\newblock
\newblock
\newblock
\shownote{https://qz.com/1367177/if-ai-is-going-to-be-the-worlds-doctor-it-needs-better-textbooks.
  Accessed: 2018-09-16.}


\bibitem[\protect\citeauthoryear{Giang}{Giang}{2018}]%
        {giang_2018}
\bibfield{author}{\bibinfo{person}{Vivian Giang}.}
  \bibinfo{year}{2018}\natexlab{}.
\newblock \bibinfo{title}{The potential hidden bias in automated hiring
  systems}.
\newblock
\newblock
\newblock
\shownote{https://www.fastcompany.com/40566971/the-potential-hidden-bias-in-automated-hiring-systems.
  Accessed: 2018-09-03.}


\bibitem[\protect\citeauthoryear{Google}{Google}{2018a}]%
        {googleDesign}
\bibfield{author}{\bibinfo{person}{Google}.} \bibinfo{year}{2018}\natexlab{a}.
\newblock \bibinfo{title}{The UX of AI - Library}.
\newblock
\newblock
\newblock
\shownote{https://design.google/ library/ux-ai/. Accessed: 2018-08-28.}


\bibitem[\protect\citeauthoryear{Google}{Google}{2018b}]%
        {google_whatif_2018}
\bibfield{author}{\bibinfo{person}{Google}.} \bibinfo{year}{2018}\natexlab{b}.
\newblock \bibinfo{title}{The What-If Tool: Code-free probing of machine
  learning models}.
\newblock
\newblock
\newblock
\shownote{https://ai.googleblog.com/2018/09/the-what-if-tool-code-free-probing-of.html.
  Accessed: 2018-09-18.}


\bibitem[\protect\citeauthoryear{Green and Hu}{Green and Hu}{2018}]%
        {greenhumyth}
\bibfield{author}{\bibinfo{person}{Ben Green} {and} \bibinfo{person}{Lily Hu}.}
  \bibinfo{year}{2018}\natexlab{}.
\newblock \showarticletitle{The myth in the methodology: Towards a
  recontextualization of fairness in machine learning}. In
  \bibinfo{booktitle}{\emph{the ICML 2018 Debates Workshop}}.
\newblock


\bibitem[\protect\citeauthoryear{Hamidi, Scheuerman, and Branham}{Hamidi
  et~al\mbox{.}}{2018}]%
        {hamidi2018gender}
\bibfield{author}{\bibinfo{person}{Foad Hamidi}, \bibinfo{person}{Morgan~Klaus
  Scheuerman}, {and} \bibinfo{person}{Stacy~M Branham}.}
  \bibinfo{year}{2018}\natexlab{}.
\newblock \showarticletitle{Gender recognition or gender reductionism?: The
  social implications of embedded gender recognition systems}. In
  \bibinfo{booktitle}{\emph{Proceedings of the 2018 CHI Conference on Human
  Factors in Computing Systems (CHI 2018)}}. ACM, \bibinfo{pages}{8}.
\newblock


\bibitem[\protect\citeauthoryear{Hanington and Martin}{Hanington and
  Martin}{2012}]%
        {hanington2012universal}
\bibfield{author}{\bibinfo{person}{Bruce Hanington} {and}
  \bibinfo{person}{Bella Martin}.} \bibinfo{year}{2012}\natexlab{}.
\newblock \bibinfo{booktitle}{\emph{Universal methods of design: 100 ways to
  research complex problems, develop innovative ideas, and design effective
  solutions}}.
\newblock \bibinfo{publisher}{Rockport Publishers}.
\newblock


\bibitem[\protect\citeauthoryear{Hardt, Price, Srebro, et~al\mbox{.}}{Hardt
  et~al\mbox{.}}{2016}]%
        {hardt2016equality}
\bibfield{author}{\bibinfo{person}{Moritz Hardt}, \bibinfo{person}{Eric Price},
  \bibinfo{person}{Nati Srebro}, {et~al\mbox{.}}}
  \bibinfo{year}{2016}\natexlab{}.
\newblock \showarticletitle{Equality of opportunity in supervised learning}. In
  \bibinfo{booktitle}{\emph{Advances in Neural Information Processing Systems
  (NeurIPS 2016)}}. \bibinfo{pages}{3315--3323}.
\newblock


\bibitem[\protect\citeauthoryear{HireVue.com}{HireVue.com}{2018}]%
        {hirevue}
\bibfield{author}{\bibinfo{person}{HireVue.com}.}
  \bibinfo{year}{2018}\natexlab{}.
\newblock \bibinfo{title}{Video interview software for recruiting \& hiring}.
\newblock
\newblock
\newblock
\shownote{https://www.hirevue.com/. Accessed: 2018-08-28.}


\bibitem[\protect\citeauthoryear{Holstein, Hong, Tegene, McLaren, and
  Aleven}{Holstein et~al\mbox{.}}{2018a}]%
        {holstein2018classroom}
\bibfield{author}{\bibinfo{person}{Kenneth Holstein}, \bibinfo{person}{Gena
  Hong}, \bibinfo{person}{Mera Tegene}, \bibinfo{person}{Bruce~M McLaren},
  {and} \bibinfo{person}{Vincent Aleven}.} \bibinfo{year}{2018}\natexlab{a}.
\newblock \showarticletitle{The classroom as a dashboard: Co-designing wearable
  cognitive augmentation for K-12 teachers}. In
  \bibinfo{booktitle}{\emph{Proceedings of the Eighth International Conference
  on Learning Analytics and Knowledge (LAK 2018)}}. ACM,
  \bibinfo{pages}{79--88}.
\newblock


\bibitem[\protect\citeauthoryear{Holstein, McLaren, and Aleven}{Holstein
  et~al\mbox{.}}{2017}]%
        {holstein2017intelligent}
\bibfield{author}{\bibinfo{person}{Kenneth Holstein}, \bibinfo{person}{Bruce~M
  McLaren}, {and} \bibinfo{person}{Vincent Aleven}.}
  \bibinfo{year}{2017}\natexlab{}.
\newblock \showarticletitle{Intelligent tutors as teachers' aides: Exploring
  teacher needs for real-time analytics in blended classrooms}. In
  \bibinfo{booktitle}{\emph{Proceedings of the Seventh International Learning
  Analytics and Knowledge Conference (LAK 2017)}}. ACM,
  \bibinfo{pages}{257--266}.
\newblock


\bibitem[\protect\citeauthoryear{Holstein, McLaren, and Aleven}{Holstein
  et~al\mbox{.}}{2018b}]%
        {holstein2018student}
\bibfield{author}{\bibinfo{person}{Kenneth Holstein}, \bibinfo{person}{Bruce~M
  McLaren}, {and} \bibinfo{person}{Vincent Aleven}.}
  \bibinfo{year}{2018}\natexlab{b}.
\newblock \showarticletitle{Student learning benefits of a mixed-reality
  teacher awareness tool in AI-enhanced classrooms}. In
  \bibinfo{booktitle}{\emph{Proceedings of the International Conference on
  Artificial Intelligence in Education (AIED 2018)}}. Springer,
  \bibinfo{pages}{154--168}.
\newblock


\bibitem[\protect\citeauthoryear{Holtzblatt and Jones}{Holtzblatt and
  Jones}{1993}]%
        {holtzblatt1993contextual}
\bibfield{author}{\bibinfo{person}{Karen Holtzblatt} {and}
  \bibinfo{person}{Sandra Jones}.} \bibinfo{year}{1993}\natexlab{}.
\newblock \showarticletitle{Contextual inquiry: A participatory technique for
  system design}.
\newblock \bibinfo{journal}{\emph{Participatory design: Principles and
  practices}} (\bibinfo{year}{1993}), \bibinfo{pages}{177--210}.
\newblock


\bibitem[\protect\citeauthoryear{Institute}{Institute}{2018}]%
        {AINow}
\bibfield{author}{\bibinfo{person}{AI~Now Institute}.}
  \bibinfo{year}{2018}\natexlab{}.
\newblock \bibinfo{title}{AI Now Institute}.
\newblock
\newblock
\newblock
\shownote{https://ainowinstitute.org. Accessed: 2018-08-03.}


\bibitem[\protect\citeauthoryear{Jain, Pecune, Matsuyama, and Cassell}{Jain
  et~al\mbox{.}}{2018}]%
        {jain2018user}
\bibfield{author}{\bibinfo{person}{Alankar Jain}, \bibinfo{person}{Florian
  Pecune}, \bibinfo{person}{Yoichi Matsuyama}, {and} \bibinfo{person}{Justine
  Cassell}.} \bibinfo{year}{2018}\natexlab{}.
\newblock \showarticletitle{A user simulator architecture for socially-aware
  conversational agents}. In \bibinfo{booktitle}{\emph{Proceedings of the 18th
  International Conference on Intelligent Virtual Agents (IVA 2018)}}. ACM,
  \bibinfo{pages}{133--140}.
\newblock


\bibitem[\protect\citeauthoryear{Janzen and Saiedian}{Janzen and
  Saiedian}{2005}]%
        {janzen2005test}
\bibfield{author}{\bibinfo{person}{David Janzen} {and} \bibinfo{person}{Hossein
  Saiedian}.} \bibinfo{year}{2005}\natexlab{}.
\newblock \showarticletitle{Test-driven development concepts, taxonomy, and
  future direction}.
\newblock \bibinfo{journal}{\emph{Computer}} \bibinfo{volume}{38},
  \bibinfo{number}{9} (\bibinfo{year}{2005}), \bibinfo{pages}{43--50}.
\newblock


\bibitem[\protect\citeauthoryear{Jia, Xu, Karanam, and Voida}{Jia
  et~al\mbox{.}}{2016}]%
        {jia2016personality}
\bibfield{author}{\bibinfo{person}{Yuan Jia}, \bibinfo{person}{Bin Xu},
  \bibinfo{person}{Yamini Karanam}, {and} \bibinfo{person}{Stephen Voida}.}
  \bibinfo{year}{2016}\natexlab{}.
\newblock \showarticletitle{Personality-targeted gamification: A survey study
  on personality traits and motivational affordances}. In
  \bibinfo{booktitle}{\emph{Proceedings of the 2016 CHI Conference on Human
  Factors in Computing Systems (CHI 2016)}}. ACM, \bibinfo{pages}{2001--2013}.
\newblock


\bibitem[\protect\citeauthoryear{Kallus and Zhou}{Kallus and Zhou}{2018}]%
        {kallus2018residual}
\bibfield{author}{\bibinfo{person}{Nathan Kallus} {and} \bibinfo{person}{Angela
  Zhou}.} \bibinfo{year}{2018}\natexlab{}.
\newblock \bibinfo{title}{Residual unfairness in fair machine learning from
  prejudiced data}.
\newblock
\newblock
\newblock
\shownote{CoRR arXiv:1806.02887.}


\bibitem[\protect\citeauthoryear{Kamar}{Kamar}{2016}]%
        {kamar2016directions}
\bibfield{author}{\bibinfo{person}{Ece Kamar}.}
  \bibinfo{year}{2016}\natexlab{}.
\newblock \showarticletitle{Directions in hybrid intelligence: Complementing AI
  systems with human intelligence.}. In \bibinfo{booktitle}{\emph{Proceedings
  of the 2016 International Joint Conference on Artificial Intelligence (IJCAI
  2016)}}. \bibinfo{pages}{4070--4073}.
\newblock


\bibitem[\protect\citeauthoryear{Kamar, Kapoor, and Horvitz}{Kamar
  et~al\mbox{.}}{2015}]%
        {kamar2015identifying}
\bibfield{author}{\bibinfo{person}{Ece Kamar}, \bibinfo{person}{Ashish Kapoor},
  {and} \bibinfo{person}{Eric Horvitz}.} \bibinfo{year}{2015}\natexlab{}.
\newblock \showarticletitle{Identifying and accounting for task-dependent bias
  in crowdsourcing}. In \bibinfo{booktitle}{\emph{Third AAAI Conference on
  Human Computation and Crowdsourcing (HCOMP 2015)}}.
\newblock


\bibitem[\protect\citeauthoryear{Kay, Matuszek, and Munson}{Kay
  et~al\mbox{.}}{2015}]%
        {kay2015unequal}
\bibfield{author}{\bibinfo{person}{Matthew Kay}, \bibinfo{person}{Cynthia
  Matuszek}, {and} \bibinfo{person}{Sean~A Munson}.}
  \bibinfo{year}{2015}\natexlab{}.
\newblock \showarticletitle{Unequal representation and gender stereotypes in
  image search results for occupations}. In
  \bibinfo{booktitle}{\emph{Proceedings of the 2015 CHI Conference on Human
  Factors in Computing Systems (CHI 2015)}}. ACM, \bibinfo{pages}{3819--3828}.
\newblock


\bibitem[\protect\citeauthoryear{Kery, Radensky, Arya, John, and Myers}{Kery
  et~al\mbox{.}}{2018}]%
        {kery2018story}
\bibfield{author}{\bibinfo{person}{Mary~Beth Kery}, \bibinfo{person}{Marissa
  Radensky}, \bibinfo{person}{Mahima Arya}, \bibinfo{person}{Bonnie~E John},
  {and} \bibinfo{person}{Brad~A Myers}.} \bibinfo{year}{2018}\natexlab{}.
\newblock \showarticletitle{The story in the notebook: Exploratory data science
  using a literate programming tool}. In \bibinfo{booktitle}{\emph{Proceedings
  of the 2018 CHI Conference on Human Factors in Computing Systems (CHI
  2018)}}. ACM, \bibinfo{pages}{174}.
\newblock


\bibitem[\protect\citeauthoryear{Kilbertus, Gasc{\'o}n, Kusner, Veale, Gummadi,
  and Weller}{Kilbertus et~al\mbox{.}}{2018}]%
        {kilbertus2018blind}
\bibfield{author}{\bibinfo{person}{Niki Kilbertus}, \bibinfo{person}{Adri{\`a}
  Gasc{\'o}n}, \bibinfo{person}{Matt~J Kusner}, \bibinfo{person}{Michael
  Veale}, \bibinfo{person}{Krishna~P Gummadi}, {and} \bibinfo{person}{Adrian
  Weller}.} \bibinfo{year}{2018}\natexlab{}.
\newblock \showarticletitle{Blind justice: Fairness with encrypted sensitive
  attributes}.
\newblock \bibinfo{journal}{\emph{Proceedings of the Thirty-Fifth International
  Conference on Machine Learning (ICML 2018)}}.
\newblock


\bibitem[\protect\citeauthoryear{Kleinberg, Mullainathan, and
  Raghavan}{Kleinberg et~al\mbox{.}}{2016}]%
        {kleinberg2016inherent}
\bibfield{author}{\bibinfo{person}{Jon Kleinberg}, \bibinfo{person}{Sendhil
  Mullainathan}, {and} \bibinfo{person}{Manish Raghavan}.}
  \bibinfo{year}{2016}\natexlab{}.
\newblock \showarticletitle{Inherent trade-offs in the fair determination of
  risk scores}.
\newblock \bibinfo{journal}{\emph{Proceedings of the Eighth Innovations in
  Theoretical Computer Science Conference (ITCS 2017)}} (\bibinfo{year}{2016}).
\newblock


\bibitem[\protect\citeauthoryear{Kulesza, Amershi, Caruana, Fisher, and
  Charles}{Kulesza et~al\mbox{.}}{2014}]%
        {kulesza2014structured}
\bibfield{author}{\bibinfo{person}{Todd Kulesza}, \bibinfo{person}{Saleema
  Amershi}, \bibinfo{person}{Rich Caruana}, \bibinfo{person}{Danyel Fisher},
  {and} \bibinfo{person}{Denis Charles}.} \bibinfo{year}{2014}\natexlab{}.
\newblock \showarticletitle{Structured labeling for facilitating concept
  evolution in machine learning}. In \bibinfo{booktitle}{\emph{Proceedings of
  the 2014 CHI Conference on Human Factors in Computing Systems (CHI 2014)}}.
  ACM, \bibinfo{pages}{3075--3084}.
\newblock


\bibitem[\protect\citeauthoryear{Kulesza, Burnett, Wong, and Stumpf}{Kulesza
  et~al\mbox{.}}{2015}]%
        {kulesza2015principles}
\bibfield{author}{\bibinfo{person}{Todd Kulesza}, \bibinfo{person}{Margaret
  Burnett}, \bibinfo{person}{Weng-Keen Wong}, {and} \bibinfo{person}{Simone
  Stumpf}.} \bibinfo{year}{2015}\natexlab{}.
\newblock \showarticletitle{Principles of explanatory debugging to personalize
  interactive machine learning}. In \bibinfo{booktitle}{\emph{Proceedings of
  the 20th International Conference on Intelligent User Interfaces (IUI
  2015)}}. ACM, \bibinfo{pages}{126--137}.
\newblock


\bibitem[\protect\citeauthoryear{Kusner, Loftus, Russell, and Silva}{Kusner
  et~al\mbox{.}}{2017}]%
        {kusner2017counterfactual}
\bibfield{author}{\bibinfo{person}{Matt~J Kusner}, \bibinfo{person}{Joshua
  Loftus}, \bibinfo{person}{Chris Russell}, {and} \bibinfo{person}{Ricardo
  Silva}.} \bibinfo{year}{2017}\natexlab{}.
\newblock \showarticletitle{Counterfactual fairness}. In
  \bibinfo{booktitle}{\emph{Advances in Neural Information Processing Systems
  (NeurIPS 2017)}}. \bibinfo{pages}{4066--4076}.
\newblock


\bibitem[\protect\citeauthoryear{Lakkaraju, Kamar, Caruana, and
  Horvitz}{Lakkaraju et~al\mbox{.}}{2017}]%
        {lakkaraju2017identifying}
\bibfield{author}{\bibinfo{person}{Himabindu Lakkaraju}, \bibinfo{person}{Ece
  Kamar}, \bibinfo{person}{Rich Caruana}, {and} \bibinfo{person}{Eric
  Horvitz}.} \bibinfo{year}{2017}\natexlab{}.
\newblock \showarticletitle{Identifying Unknown Unknowns in the Open World:
  Representations and Policies for Guided Exploration.}. In
  \bibinfo{booktitle}{\emph{Proceedings of the AAAI Conference on Artificial
  Intelligence (AAAI 2017)}}.
\newblock


\bibitem[\protect\citeauthoryear{Larson, Mattu, Kirchner, and Angwin}{Larson
  et~al\mbox{.}}{2016}]%
        {larson2016we}
\bibfield{author}{\bibinfo{person}{Jeff Larson}, \bibinfo{person}{Surya Mattu},
  \bibinfo{person}{Lauren Kirchner}, {and} \bibinfo{person}{Julia Angwin}.}
  \bibinfo{year}{2016}\natexlab{}.
\newblock \showarticletitle{How we analyzed the COMPAS recidivism algorithm}.
\newblock \bibinfo{journal}{\emph{ProPublica (5 2016)}}  \bibinfo{volume}{9}
  (\bibinfo{year}{2016}).
\newblock


\bibitem[\protect\citeauthoryear{Lee}{Lee}{2018}]%
        {lee2018understanding}
\bibfield{author}{\bibinfo{person}{Min~Kyung Lee}.}
  \bibinfo{year}{2018}\natexlab{}.
\newblock \showarticletitle{Understanding perception of algorithmic decisions:
  Fairness, trust, and emotion in response to algorithmic management}.
\newblock \bibinfo{journal}{\emph{Big Data \& Society}} \bibinfo{volume}{5},
  \bibinfo{number}{1} (\bibinfo{year}{2018}),
  \bibinfo{pages}{2053951718756684}.
\newblock


\bibitem[\protect\citeauthoryear{Lee and Baykal}{Lee and Baykal}{2017}]%
        {lee2017algorithmic}
\bibfield{author}{\bibinfo{person}{Min~Kyung Lee} {and} \bibinfo{person}{Su
  Baykal}.} \bibinfo{year}{2017}\natexlab{}.
\newblock \showarticletitle{Algorithmic mediation in group decisions: Fairness
  perceptions of algorithmically mediated vs. discussion-based social
  division}. In \bibinfo{booktitle}{\emph{Proceedings of the 2017 ACM
  Conference on Computer Supported Cooperative Work (CSCW 2017)}}.
  \bibinfo{pages}{1035--1048}.
\newblock


\bibitem[\protect\citeauthoryear{Liu, Reyzin, and Ziebart}{Liu
  et~al\mbox{.}}{2015}]%
        {liu2015shift}
\bibfield{author}{\bibinfo{person}{Anqi Liu}, \bibinfo{person}{Lev Reyzin},
  {and} \bibinfo{person}{Brian~D Ziebart}.} \bibinfo{year}{2015}\natexlab{}.
\newblock \showarticletitle{Shift-pessimistic active learning using robust
  bias-aware prediction}. In \bibinfo{booktitle}{\emph{Proceedings of the AAAI
  Conference on Artificial Intelligence (AAAI 2015)}}.
  \bibinfo{pages}{2764--2770}.
\newblock


\bibitem[\protect\citeauthoryear{Liu and Singh}{Liu and Singh}{2002}]%
        {liu2002makebelieve}
\bibfield{author}{\bibinfo{person}{Hugo Liu} {and} \bibinfo{person}{Push
  Singh}.} \bibinfo{year}{2002}\natexlab{}.
\newblock \showarticletitle{MAKEBELIEVE: Using commonsense knowledge to
  generate stories}. In \bibinfo{booktitle}{\emph{Proceedings of the Fourteenth
  Innovative Applications of Artificial Intelligence Conference (IAAI 2002)}}.
  \bibinfo{pages}{957--958}.
\newblock


\bibitem[\protect\citeauthoryear{Liu, Dean, Rolf, Simchowitz, and Hardt}{Liu
  et~al\mbox{.}}{2018}]%
        {liu2018delayed}
\bibfield{author}{\bibinfo{person}{Lydia~T Liu}, \bibinfo{person}{Sarah Dean},
  \bibinfo{person}{Esther Rolf}, \bibinfo{person}{Max Simchowitz}, {and}
  \bibinfo{person}{Moritz Hardt}.} \bibinfo{year}{2018}\natexlab{}.
\newblock \showarticletitle{Delayed impact of fair machine learning}.
\newblock \bibinfo{journal}{\emph{Proceedings of the Thirty-fifth International
  Conference on Machine Learning (ICML 2018)}} (\bibinfo{year}{2018}).
\newblock


\bibitem[\protect\citeauthoryear{Lomas}{Lomas}{2018a}]%
        {lomas_2018}
\bibfield{author}{\bibinfo{person}{Natasha Lomas}.}
  \bibinfo{year}{2018}\natexlab{a}.
\newblock \bibinfo{title}{Accenture wants to beat unfair AI with a professional
  toolkit}.
\newblock
\newblock
\newblock
\shownote{https://techcrun
  ch.com/2018/06/09/accenture-wants-to-beat-unfair-ai-with-a-professional-toolkit/.
  Accessed: 2018-06-14.}


\bibitem[\protect\citeauthoryear{Lomas}{Lomas}{2018b}]%
        {lomas_2018_IBM}
\bibfield{author}{\bibinfo{person}{Natasha Lomas}.}
  \bibinfo{year}{2018}\natexlab{b}.
\newblock \bibinfo{title}{IBM launches cloud tool to detect AI bias and explain
  automated decisions}.
\newblock
\newblock
\newblock
\shownote{https://techcrunch.com/2018/09/19/ibm-launches-cloud-tool-to-detect-ai-bias-and-explain-automated-decisions.
  Accessed: 2018-09-20.}


\bibitem[\protect\citeauthoryear{Lum and Isaac}{Lum and Isaac}{2016}]%
        {lum2016predict}
\bibfield{author}{\bibinfo{person}{Kristian Lum} {and} \bibinfo{person}{William
  Isaac}.} \bibinfo{year}{2016}\natexlab{}.
\newblock \showarticletitle{To predict and serve?}
\newblock \bibinfo{journal}{\emph{Significance}} \bibinfo{volume}{13},
  \bibinfo{number}{5} (\bibinfo{year}{2016}), \bibinfo{pages}{14--19}.
\newblock


\bibitem[\protect\citeauthoryear{Lyu, Kantardzic, and Sethi}{Lyu
  et~al\mbox{.}}{2018}]%
        {lyu2018sloppiness}
\bibfield{author}{\bibinfo{person}{Lingyu Lyu}, \bibinfo{person}{Mehmed
  Kantardzic}, {and} \bibinfo{person}{Tegjyot~Singh Sethi}.}
  \bibinfo{year}{2018}\natexlab{}.
\newblock \showarticletitle{Sloppiness mitigation in crowdsourcing: detecting
  and correcting bias for crowd scoring tasks}.
\newblock \bibinfo{journal}{\emph{International Journal of Data Science and
  Analytics}} (\bibinfo{year}{2018}), \bibinfo{pages}{1--21}.
\newblock


\bibitem[\protect\citeauthoryear{Maclellan, Harpstead, Patel, and
  Koedinger}{Maclellan et~al\mbox{.}}{2016}]%
        {maclellan2016apprentice}
\bibfield{author}{\bibinfo{person}{Christopher~J Maclellan},
  \bibinfo{person}{Erik Harpstead}, \bibinfo{person}{Rony Patel}, {and}
  \bibinfo{person}{Kenneth~R Koedinger}.} \bibinfo{year}{2016}\natexlab{}.
\newblock \showarticletitle{The Apprentice Learner architecture: Closing the
  loop between learning theory and educational data.}. In
  \bibinfo{booktitle}{\emph{Proceedings of the 2016 International Conference on
  Educational Data Mining (EDM 2016)}}. \bibinfo{pages}{151--158}.
\newblock


\bibitem[\protect\citeauthoryear{Malouff and Thorsteinsson}{Malouff and
  Thorsteinsson}{2016}]%
        {malouff2016bias}
\bibfield{author}{\bibinfo{person}{John~M Malouff} {and}
  \bibinfo{person}{Einar~B Thorsteinsson}.} \bibinfo{year}{2016}\natexlab{}.
\newblock \showarticletitle{Bias in grading: A meta-analysis of experimental
  research findings}.
\newblock \bibinfo{journal}{\emph{Australian Journal of Education}}
  \bibinfo{volume}{60}, \bibinfo{number}{3} (\bibinfo{year}{2016}),
  \bibinfo{pages}{245--256}.
\newblock


\bibitem[\protect\citeauthoryear{MURAL}{MURAL}{2018}]%
        {mural}
\bibfield{author}{\bibinfo{person}{MURAL}.} \bibinfo{year}{2018}\natexlab{}.
\newblock \bibinfo{title}{MURAL - Make remote design work.}
\newblock
\newblock
\newblock
\shownote{https://mural.co/. Accessed: 2018-08-02.}


\bibitem[\protect\citeauthoryear{Narayanan}{Narayanan}{2018}]%
        {21definitions}
\bibfield{author}{\bibinfo{person}{Arvind Narayanan}.}
  \bibinfo{year}{2018}\natexlab{}.
\newblock \showarticletitle{21 fairness definitions and their politics}.
\newblock \bibinfo{journal}{\emph{FAT* 2018 tutorial}} (\bibinfo{year}{2018}).
\newblock


\bibitem[\protect\citeauthoryear{Noble}{Noble}{2018}]%
        {noble2018algorithms}
\bibfield{author}{\bibinfo{person}{Safiya~Umoja Noble}.}
  \bibinfo{year}{2018}\natexlab{}.
\newblock \bibinfo{booktitle}{\emph{Algorithms of oppression: How search
  engines reinforce racism}}.
\newblock \bibinfo{publisher}{NYU Press}.
\newblock


\bibitem[\protect\citeauthoryear{Nushi, Kamar, Horvitz, and Kossmann}{Nushi
  et~al\mbox{.}}{2017}]%
        {nushi2017human}
\bibfield{author}{\bibinfo{person}{Besmira Nushi}, \bibinfo{person}{Ece Kamar},
  \bibinfo{person}{Eric Horvitz}, {and} \bibinfo{person}{Donald Kossmann}.}
  \bibinfo{year}{2017}\natexlab{}.
\newblock \showarticletitle{On human intellect and machine failures:
  Troubleshooting integrative machine learning systems}. In
  \bibinfo{booktitle}{\emph{Proceedings of the AAAI Conference on Artificial
  Intelligence (AAAI 2017)}}. \bibinfo{pages}{1017--1025}.
\newblock


\bibitem[\protect\citeauthoryear{of~Education~(ED)}{of~Education~(ED)}{2018}]%
        {home_2018}
\bibfield{author}{\bibinfo{person}{US~Department of Education~(ED)}.}
  \bibinfo{year}{2018}\natexlab{}.
\newblock \bibinfo{title}{Family Educational Rights and Privacy Act (FERPA)}.
\newblock
\newblock
\newblock
\shownote{https://www2.ed.gov/policy/gen/guid/fpco/ferpa/index.html. Accessed:
  2018-09-04.}


\bibitem[\protect\citeauthoryear{on~AI}{on~AI}{2018}]%
        {partnershipOnAI}
\bibfield{author}{\bibinfo{person}{Partnership on AI}.}
  \bibinfo{year}{2018}\natexlab{}.
\newblock \bibinfo{title}{The Partnership on AI}.
\newblock
\newblock
\newblock
\shownote{https://www.partnershiponai.org. Accessed: 2018-09-03.}


\bibitem[\protect\citeauthoryear{Powles and Hodson}{Powles and Hodson}{2017}]%
        {powles2017google}
\bibfield{author}{\bibinfo{person}{Julia Powles} {and} \bibinfo{person}{Hal
  Hodson}.} \bibinfo{year}{2017}\natexlab{}.
\newblock \showarticletitle{Google DeepMind and healthcare in an age of
  algorithms}.
\newblock \bibinfo{journal}{\emph{Health and Technology}} \bibinfo{volume}{7},
  \bibinfo{number}{4} (\bibinfo{year}{2017}), \bibinfo{pages}{351--367}.
\newblock


\bibitem[\protect\citeauthoryear{pymetrics}{pymetrics}{2018}]%
        {pymetrics}
\bibfield{author}{\bibinfo{person}{pymetrics}.}
  \bibinfo{year}{2018}\natexlab{}.
\newblock \bibinfo{title}{matching talent to opportunity}.
\newblock
\newblock
\newblock
\shownote{https://www.pymetrics.com/. Accessed: 2018-08-28.}


\bibitem[\protect\citeauthoryear{Qualtrics}{Qualtrics}{2013}]%
        {qualtrics2013qualtrics}
\bibfield{author}{\bibinfo{person}{Qualtrics}.}
  \bibinfo{year}{2013}\natexlab{}.
\newblock \showarticletitle{Qualtrics}.
\newblock \bibinfo{journal}{\emph{Provo, UT, USA}} (\bibinfo{year}{2013}).
\newblock


\bibitem[\protect\citeauthoryear{Rader and Gray}{Rader and Gray}{2015}]%
        {rader2015understanding}
\bibfield{author}{\bibinfo{person}{Emilee Rader} {and} \bibinfo{person}{Rebecca
  Gray}.} \bibinfo{year}{2015}\natexlab{}.
\newblock \showarticletitle{Understanding user beliefs about algorithmic
  curation in the Facebook news feed}. In \bibinfo{booktitle}{\emph{Proceedings
  of the 2015 CHI Conference on Human Factors in Computing Systems (CHI
  2015)}}. ACM, \bibinfo{pages}{173--182}.
\newblock


\bibitem[\protect\citeauthoryear{Raghavan, Slivkins, Vaughan, and Wu}{Raghavan
  et~al\mbox{.}}{2018}]%
        {raghavan2018externalities}
\bibfield{author}{\bibinfo{person}{Manish Raghavan},
  \bibinfo{person}{Aleksandrs Slivkins}, \bibinfo{person}{Jennifer~Wortman
  Vaughan}, {and} \bibinfo{person}{Zhiwei~Steven Wu}.}
  \bibinfo{year}{2018}\natexlab{}.
\newblock \showarticletitle{The externalities of exploration and how data
  diversity helps exploitation}. In \bibinfo{booktitle}{\emph{Proceedings of
  the Thirty-first Annual Conference on Learning Theory (COLT 2018)}}.
\newblock


\bibitem[\protect\citeauthoryear{Reisman, Schultz, Crawford, and
  Whittaker}{Reisman et~al\mbox{.}}{2018}]%
        {reisman2018algorithmic}
\bibfield{author}{\bibinfo{person}{Dillon Reisman}, \bibinfo{person}{Jason
  Schultz}, \bibinfo{person}{K Crawford}, {and} \bibinfo{person}{M Whittaker}.}
  \bibinfo{year}{2018}\natexlab{}.
\newblock \showarticletitle{Algorithmic impact assessments: A practical
  framework for public agency accountability}.
\newblock \bibinfo{journal}{\emph{AI Now Institute}} (\bibinfo{year}{2018}).
\newblock


\bibitem[\protect\citeauthoryear{Schlesinger, O'Hara, and Taylor}{Schlesinger
  et~al\mbox{.}}{2018}]%
        {schlesinger2018let}
\bibfield{author}{\bibinfo{person}{Ari Schlesinger}, \bibinfo{person}{Kenton~P
  O'Hara}, {and} \bibinfo{person}{Alex~S Taylor}.}
  \bibinfo{year}{2018}\natexlab{}.
\newblock \showarticletitle{Let's talk about race: Identity, chatbots, and AI}.
  In \bibinfo{booktitle}{\emph{Proceedings of the 2018 CHI Conference on Human
  Factors in Computing Systems (CHI 2018)}}. ACM, \bibinfo{pages}{315}.
\newblock


\bibitem[\protect\citeauthoryear{Sculley, Holt, Golovin, Davydov, Phillips,
  Ebner, Chaudhary, Young, Crespo, and Dennison}{Sculley et~al\mbox{.}}{2015}]%
        {sculley2015hidden}
\bibfield{author}{\bibinfo{person}{D. Sculley}, \bibinfo{person}{Gary Holt},
  \bibinfo{person}{Daniel Golovin}, \bibinfo{person}{Eugene Davydov},
  \bibinfo{person}{Todd Phillips}, \bibinfo{person}{Dietmar Ebner},
  \bibinfo{person}{Vinay Chaudhary}, \bibinfo{person}{Michael Young},
  \bibinfo{person}{Jean-Francois Crespo}, {and} \bibinfo{person}{Dan
  Dennison}.} \bibinfo{year}{2015}\natexlab{}.
\newblock \showarticletitle{Hidden technical debt in machine learning systems}.
  In \bibinfo{booktitle}{\emph{Advances in Neural Information Processing
  Systems (NeurIPS 2015)}}. \bibinfo{pages}{2503--2511}.
\newblock


\bibitem[\protect\citeauthoryear{Selbst, danah boyd, Friedler,
  Venkatasubramanian, and Vertesi}{Selbst et~al\mbox{.}}{2019}]%
        {selbst2018fairness}
\bibfield{author}{\bibinfo{person}{Andrew~D Selbst}, \bibinfo{person}{danah
  boyd}, \bibinfo{person}{Sorelle Friedler}, \bibinfo{person}{Suresh
  Venkatasubramanian}, {and} \bibinfo{person}{Janet Vertesi}.}
  \bibinfo{year}{2019}\natexlab{}.
\newblock \showarticletitle{Fairness and abstraction in sociotechnical
  systems}. In \bibinfo{booktitle}{\emph{ACM Conference on Fairness,
  Accountability, and Transparency (FAT* 2018)}}.
\newblock


\bibitem[\protect\citeauthoryear{Society}{Society}{2018}]%
        {AAprimer}
\bibfield{author}{\bibinfo{person}{Data~\& Society}.}
  \bibinfo{year}{2018}\natexlab{}.
\newblock \showarticletitle{Algorithmic accountability: A primer}.
\newblock \bibinfo{journal}{\emph{Data \& Society}} (\bibinfo{year}{2018}).
\newblock


\bibitem[\protect\citeauthoryear{Springer, Garcia-Gathright, and
  Cramer}{Springer et~al\mbox{.}}{2018}]%
        {springerGarciaCramer2018}
\bibfield{author}{\bibinfo{person}{Aaron Springer}, \bibinfo{person}{J.
  Garcia-Gathright}, {and} \bibinfo{person}{Henriette Cramer}.}
  \bibinfo{year}{2018}\natexlab{}.
\newblock \showarticletitle{Assessing and addressing algorithmic bias---But
  before we get there}. In \bibinfo{booktitle}{\emph{Proceedings of the AAAI
  2018 Spring Symposium: Designing the User Experience of Artificial
  Intelligence}}.
\newblock


\bibitem[\protect\citeauthoryear{Stokes}{Stokes}{1997}]%
        {stokes2011pasteur}
\bibfield{author}{\bibinfo{person}{Donald~E Stokes}.}
  \bibinfo{year}{1997}\natexlab{}.
\newblock \bibinfo{booktitle}{\emph{Pasteur's quadrant: Basic science and
  technological innovation}}.
\newblock \bibinfo{publisher}{Brookings Institution Press}.
\newblock


\bibitem[\protect\citeauthoryear{Tan, Adebayo, Inkpen, and Kamar}{Tan
  et~al\mbox{.}}{2018}]%
        {tan2018investigating}
\bibfield{author}{\bibinfo{person}{Sarah Tan}, \bibinfo{person}{Julius
  Adebayo}, \bibinfo{person}{Kori Inkpen}, {and} \bibinfo{person}{Ece Kamar}.}
  \bibinfo{year}{2018}\natexlab{}.
\newblock \bibinfo{title}{Investigating Human+Machine Complementarity for
  Recidivism Predictions}.
\newblock
\newblock
\newblock
\shownote{CoRR arXiv:1808.09123.}


\bibitem[\protect\citeauthoryear{Thubron}{Thubron}{2018}]%
        {thubron_2018}
\bibfield{author}{\bibinfo{person}{Rob Thubron}.}
  \bibinfo{year}{2018}\natexlab{}.
\newblock \bibinfo{title}{IBM secretly used NYPD CCTV footage to train its
  facial recognition systems}.
\newblock
\newblock
\newblock
\shownote{https://www.techspot.com/news/76323-ibm-secretly-used-nypd-cctv-footage-train-facial.html.
  Accessed: 2018-09-16.}


\bibitem[\protect\citeauthoryear{Toyama}{Toyama}{2018}]%
        {toyama2018needs}
\bibfield{author}{\bibinfo{person}{Kentaro Toyama}.}
  \bibinfo{year}{2018}\natexlab{}.
\newblock \showarticletitle{From needs to aspirations in information technology
  for development}.
\newblock \bibinfo{journal}{\emph{Information Technology for Development}}
  \bibinfo{volume}{24}, \bibinfo{number}{1}, \bibinfo{pages}{15--36}.
\newblock


\bibitem[\protect\citeauthoryear{Valentine, Retelny, To, Rahmati, Doshi, and
  Bernstein}{Valentine et~al\mbox{.}}{2017}]%
        {valentine2017flash}
\bibfield{author}{\bibinfo{person}{Melissa~A Valentine},
  \bibinfo{person}{Daniela Retelny}, \bibinfo{person}{Alexandra To},
  \bibinfo{person}{Negar Rahmati}, \bibinfo{person}{Tulsee Doshi}, {and}
  \bibinfo{person}{Michael~S Bernstein}.} \bibinfo{year}{2017}\natexlab{}.
\newblock \showarticletitle{Flash organizations: Crowdsourcing complex work by
  structuring crowds as organizations}. In
  \bibinfo{booktitle}{\emph{Proceedings of the 2017 CHI Conference on Human
  Factors in Computing Systems (CHI 2017)}}. ACM, \bibinfo{pages}{3523--3537}.
\newblock


\bibitem[\protect\citeauthoryear{Vaughan}{Vaughan}{2018}]%
        {vaughan2017making}
\bibfield{author}{\bibinfo{person}{Jennifer~Wortman Vaughan}.}
  \bibinfo{year}{2018}\natexlab{}.
\newblock \showarticletitle{Making better use of the crowd}.
\newblock \bibinfo{journal}{\emph{Journal of Machine Learning Research}}
  \bibinfo{volume}{18}, \bibinfo{number}{193} (\bibinfo{year}{2018}),
  \bibinfo{pages}{1--46}.
\newblock


\bibitem[\protect\citeauthoryear{Veale and Binns}{Veale and Binns}{2017}]%
        {veale2017fairer}
\bibfield{author}{\bibinfo{person}{Michael Veale} {and} \bibinfo{person}{Reuben
  Binns}.} \bibinfo{year}{2017}\natexlab{}.
\newblock \showarticletitle{Fairer machine learning in the real world:
  Mitigating discrimination without collecting sensitive data}.
\newblock \bibinfo{journal}{\emph{Big Data \& Society}} \bibinfo{volume}{4},
  \bibinfo{number}{2} (\bibinfo{year}{2017}),
  \bibinfo{pages}{2053951717743530}.
\newblock


\bibitem[\protect\citeauthoryear{Veale, Van~Kleek, and Binns}{Veale
  et~al\mbox{.}}{2018}]%
        {veale2018fairness}
\bibfield{author}{\bibinfo{person}{Michael Veale}, \bibinfo{person}{Max
  Van~Kleek}, {and} \bibinfo{person}{Reuben Binns}.}
  \bibinfo{year}{2018}\natexlab{}.
\newblock \showarticletitle{Fairness and accountability design needs for
  algorithmic support in high-stakes public sector decision-making}. In
  \bibinfo{booktitle}{\emph{Proceedings of the 2018 CHI Conference on Human
  Factors in Computing Systems (CHI 2018)}}. ACM, \bibinfo{pages}{440}.
\newblock


\bibitem[\protect\citeauthoryear{Wachter-Boettcher}{Wachter-Boettcher}{2017}]%
        {wachter-boettcher_2017}
\bibfield{author}{\bibinfo{person}{Sara Wachter-Boettcher}.}
  \bibinfo{year}{2017}\natexlab{}.
\newblock \bibinfo{title}{AI recruiting tools do not eliminate bias}.
\newblock
\newblock
\newblock
\shownote{http://time.com/4993431/ai-recruiting-tools-do-not-eliminate-bias.
  Accessed: 2018-09-01.}


\bibitem[\protect\citeauthoryear{Woodruff, Fox, Rousso-Schindler, and
  Warshaw}{Woodruff et~al\mbox{.}}{2018}]%
        {woodruff2018qualitative}
\bibfield{author}{\bibinfo{person}{Allison Woodruff}, \bibinfo{person}{Sarah~E
  Fox}, \bibinfo{person}{Steven Rousso-Schindler}, {and}
  \bibinfo{person}{Jeffrey Warshaw}.} \bibinfo{year}{2018}\natexlab{}.
\newblock \showarticletitle{A qualitative exploration of perceptions of
  algorithmic fairness}. In \bibinfo{booktitle}{\emph{Proceedings of the 2018
  CHI Conference on Human Factors in Computing Systems (CHI 2018)}}. ACM,
  \bibinfo{pages}{656}.
\newblock


\bibitem[\protect\citeauthoryear{Yang}{Yang}{2018}]%
        {yang2018machine}
\bibfield{author}{\bibinfo{person}{Qian Yang}.}
  \bibinfo{year}{2018}\natexlab{}.
\newblock \showarticletitle{Machine learning as a UX design material: How can
  we imagine beyond automation, recommenders, and reminders?}. In
  \bibinfo{booktitle}{\emph{Proceedings of the AAAI 2018 Spring Symposium:
  Designing the User Experience of Artificial Intelligence}}.
\newblock


\bibitem[\protect\citeauthoryear{Yang, Suh, Chen, and Ramos}{Yang
  et~al\mbox{.}}{2018}]%
        {yang2018grounding}
\bibfield{author}{\bibinfo{person}{Qian Yang}, \bibinfo{person}{Jina Suh},
  \bibinfo{person}{Nan-Chen Chen}, {and} \bibinfo{person}{Gonzalo Ramos}.}
  \bibinfo{year}{2018}\natexlab{}.
\newblock \showarticletitle{Grounding interactive machine learning tool design
  in how non-experts actually build models}. In
  \bibinfo{booktitle}{\emph{Proceedings of the 2018 Conference on Designing
  Interactive Systems (DIS 2018)}}. ACM, \bibinfo{pages}{573--584}.
\newblock


\bibitem[\protect\citeauthoryear{Yang, Zimmerman, Steinfeld, Carey, and
  Antaki}{Yang et~al\mbox{.}}{2016}]%
        {yang2016investigating}
\bibfield{author}{\bibinfo{person}{Qian Yang}, \bibinfo{person}{John
  Zimmerman}, \bibinfo{person}{Aaron Steinfeld}, \bibinfo{person}{Lisa Carey},
  {and} \bibinfo{person}{James~F Antaki}.} \bibinfo{year}{2016}\natexlab{}.
\newblock \showarticletitle{Investigating the heart pump implant decision
  process: Opportunities for decision support tools to help}. In
  \bibinfo{booktitle}{\emph{Proceedings of the 2016 CHI Conference on Human
  Factors in Computing Systems (CHI 2016)}}. ACM, \bibinfo{pages}{4477--4488}.
\newblock


\bibitem[\protect\citeauthoryear{Zhang}{Zhang}{2015}]%
        {zhang_2015}
\bibfield{author}{\bibinfo{person}{Maggie Zhang}.}
  \bibinfo{year}{2015}\natexlab{}.
\newblock \bibinfo{title}{Google photos tags two African-Americans as gorillas
  through facial recognition software}.
\newblock
\newblock
\newblock
\shownote{https://tinyurl.com/Forbes-2015-07-01. Accessed: 2018-07-12.}


\bibitem[\protect\citeauthoryear{Zhao, Madaio, Pecune, Matsuyama, and
  Cassell}{Zhao et~al\mbox{.}}{2018}]%
        {zhao2018socially}
\bibfield{author}{\bibinfo{person}{Zian Zhao}, \bibinfo{person}{Michael
  Madaio}, \bibinfo{person}{Florian Pecune}, \bibinfo{person}{Yoichi
  Matsuyama}, {and} \bibinfo{person}{Justine Cassell}.}
  \bibinfo{year}{2018}\natexlab{}.
\newblock \showarticletitle{Socially-conditioned task reasoning for a virtual
  tutoring agent}. In \bibinfo{booktitle}{\emph{Proceedings of the Seventeenth
  International Conference on Autonomous Agents and Multiagent Systems (AAMAS
  2018)}}. \bibinfo{pages}{2265--2267}.
\newblock


\end{thebibliography}

\end{document}